\documentclass{aa}  

\usepackage{graphicx}
\usepackage{txfonts}
\usepackage{mathtools}
\usepackage{gensymb}
\usepackage{natbib}
\usepackage{array,booktabs}  
\begin{document} 

   \title{Non-thermal emission in hyper-velocity and semi-relativistic stars}
    \authorrunning{J.~R. Martinez et al.}


   \author{J.~R. Martinez\inst{1,2}, S. del Palacio\inst{2}, V. Bosch-Ramon\inst{2,3} \& G.~E. Romero\inst{2}
          }

   \institute{Facultad de Ciencias Exactas, UNLP,
              Calle 47 y 115, CP(1900), La Plata, Buenos Aires, Argentina.\\
              \email{jmartinez@iar.unlp.edu.ar}
         \and
             Instituto Argentino de Radioastronom\'ia (CCT La Plata, CONICET), C.C.5, (1894) Villa Elisa, Buenos Aires, Argentina.\\
             \email{sdelpalacio@iar.unlp.edu.ar}
        \and
            Departament de F\'isica Qu\'antica i Astrof\'isica, Institut de Ci\'encies del Cosmos (ICCUB), Universitat de Barcelona (IEEC-UB), Mart\'i i Franqu\`es 1, E-08028 Barcelona, Spain.\\
            \email{vbosch@am.ub.es}
             }

   \date{}

 
  \abstract
   {There is a population of runaway stars that move at extremely high speeds with respect to their surroundings. The fast motion and the stellar wind of these stars, plus the wind-medium interaction, can lead to particle acceleration and non-thermal radiation.}
   {We characterise the interaction between the winds of fast runaway stars and their environment, in particular to establish their potential as cosmic-ray accelerators and non-thermal emitters.}
   {We model the hydrodynamics of the interaction between the stellar wind and the surrounding material. We self-consistently calculate the injection and transport of relativistic particles in the bow shock using a multi-zone code, and compute their broadband emission from radio to $\gamma$-rays.}
   {Both the forward and reverse shocks are favourable sites for particle acceleration, although the radiative efficiency of particles is low and therefore the expected fluxes are in general rather faint.}
   {We show that high-sensitivity observations in the radio band can be used to detect the non-thermal radiation associated with bow shocks from hypervelocity and semi-relativistic stars. Hypervelocity stars are expected to be modest sources of sub-TeV cosmic rays, accounting perhaps for a $\sim 0.1$\% of that of galactic cosmic rays.}


   \maketitle
%

\section{Introduction}\label{Sec:Introduction}

Massive stars have intense ultraviolet (UV) radiation fields that accelerate the surface material, launching powerful supersonic winds. These stellar winds interact with the interstellar medium (ISM) generating two shock fronts: a forward shock (FS) that propagates through the ISM and a reverse shock (RS) that propagates through the stellar wind \citep{Weaver1977}. These shocks are potential sites for non-thermal phenomena; they have been detected in a few occasions \citep{Prajapati_2019, Sanchez-Ayaso_2018}, and have been suggested to produce galactic cosmic rays up to PeV energies \citep[e.g.][]{Aharonian2019, Morlino2021}. 

Stars that have a supersonic velocity with respect to the local ISM are known as runaway stars. 
In this scenario, the interaction region becomes bow-shaped \citep{vanBuren1988}, and thus the whole interaction structure is usually called bow shock (BS). The FS compresses and heats up dust and gas that emit mostly infrared (IR) and optical radiation \citep{Peri_2012, Kobulnicky2016}, but BSs can also accelerate particles up to relativistic energies via diffusive shock acceleration (DSA). These particles, in turn, can interact with matter and electromagnetic fields, producing broadband non-thermal (NT) radiation \citep[as predicted by, e.g.,][]{del_Valle_2012, del_Palacio_2018, de_Valle_2018}. Nonetheless, despite more than 700 BSs have been cataloged, NT emission has been clearly detected only in one of them, BD+43°3654 \citep{Benaglia_2010, Benaglia_2021}. We note that it has also been detected radio emission from the BS produced by the high-mass X-ray binary Vela X-1, but in this case the nature (thermal or NT) of the emission is still uncertain \citep{van_den_Eijnden2021}.

Hyper-velocity stars are the subclass of runaway stars with velocities of hundreds to a few thousands of km\;s$^{-1}$ \citep{Brown_2015}. Recent observational campaigns have catalogued hundreds of HVSs with the data provided by \textit{Gaia}, and models predict an ejection rate of HVSs of 10$^{-4}$--10$^{-5}$~yr$^{-1}$ \citep{Zhang_2013}.
In the case of massive HVSs, they are expected to generate strong BSs, being promissory NT sources. 
The Hills mechanism \citep{Hills_1988} explains the HVSs origin via a 3-body exchange between a stellar binary and a supermassive black hole. The black hole disrupts the binary, ejecting one of its components at great velocities. The velocity of ejection depends on the supermassive black hole mass, and the total mass and semimajor axis of the binary. Recently, this mechanism gained great support by the discovery of a $\sim 1700~{\rm km\;s^{-1}}$ A-type star ejected from Sgr A$^*$ \citep{Koposov_2020}.

\cite{Tutukov_2009} predicted a putative subclass of HVSs with semi-relativistic velocities, called semi-relativistic stars (SRSs). Numerical simulations support this prediction \citep{Loeb_2016}, and according to \cite{Dremova_2017} a modified Hills mechanism can explain their origin. This mechanism consists of the gravitational interaction of two supermassive black holes that eject stars located in their central clusters.
This mechanism predicts velocities of tens of thousands of km\,s$^{-1}$, the maximum speed of ejection of the SRSs being determined by the mass of the secondary black hole and the mass of the ejected star \citep[e.g.][]{Guillochon_2015}.

In this work, we aim to characterise the NT particle production and associated emission spectra of BSs produced by massive stars that propagate with extreme velocities. In particular, we focus on massive HVSs and a putative SRS. 

The paper is organised as follows. In Sec.~\ref{sec:Modelo}, we present a multi-zone emission model that is suitable for extreme velocity stars for which the FS is also relevant. 
We present and discuss our results in Sec.~\ref{sec:results}, and finally, we conclude with a summary of the main findings of our work in Sec.~\ref{sec:conclusions}.


\section{Model}\label{sec:Modelo}

\begin{figure*}[ht]
    \centering
    \includegraphics[ width=0.9\linewidth]{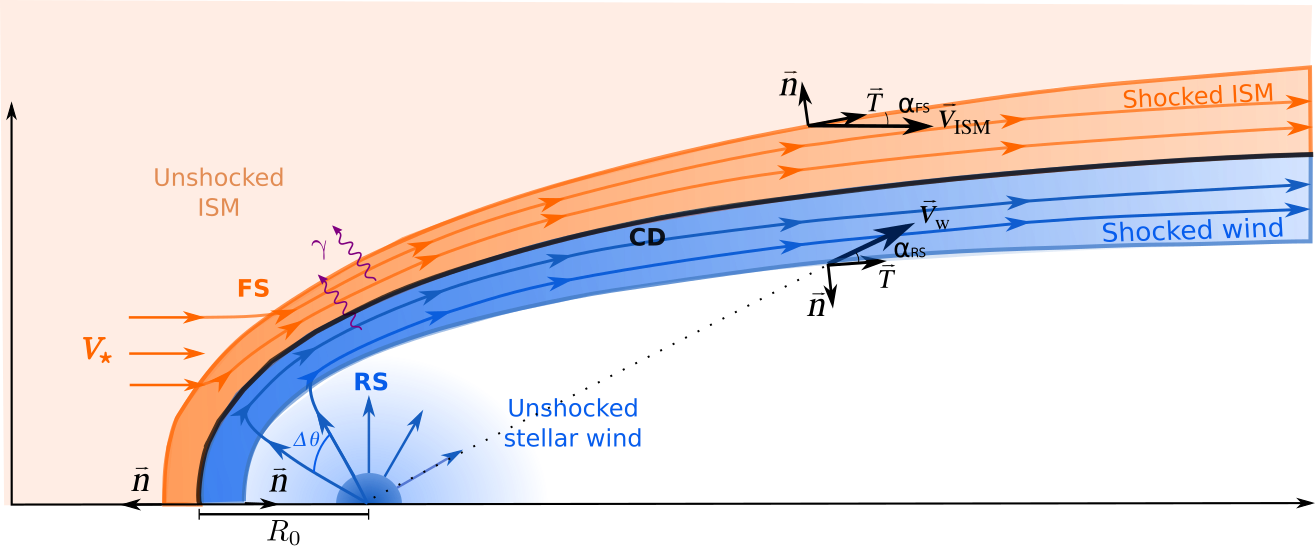}
    \caption{Sketch of the model considered. The position of the CD is represented by a black solid line, while the orange and blue regions represent the FS and the RS, respectively. The solid lines with arrows represent different streamlines in each shock, injected in different positions separated by an angle $\Delta \theta$. We also show the orientation of perpendicular and tangential vectors to the shocks in different positions, alongside $\vec{v}_{\rm w}$ and $\vec{V}_{\rm ISM} = -\vec{V}_\star$, and the angle $\alpha$ between them. Adapted from \cite{del_Palacio_2018}.}
    \label{fig:BS}
\end{figure*}

\subsection{Scenarios studied}\label{subsec:sistemas}

We define the fiducial cases to study keeping a compromise between the potential detectability of the sources and the feasibility of finding such objects. Regarding the luminosity of a BS produced by a massive star, the most important parameters are those associated with the properties of the stellar wind, and how they relate to the properties of the medium \citep{del_Palacio_2018}. We focus here particularly on massive stars in the main sequence, as this is the evolutionary stage in which they spend most of their life. These stars produce more powerful winds (and therefore more luminous BSs) for younger spectral types, although younger and more massive stars are less numerous \citep{Salpeter_1995}. 
In addition to stellar mass dependence, the energetics and radiation efficiency of the BS can also increase with the stellar spatial velocity \citep{Martinez2021}. 

Observations with the \textit{Gaia} satellite have detected several B-type stars with speeds over 100~km\,s$^{-1}$, but not a substantial quantity of O-type stars with those velocities \citep{kreuzer_2020}. According to \cite{Marchetti_2019}, the highest velocities found in HVSs are of the order of $V_\star \sim 10\,000$~km\,s$^{-1}$, although estimations above 1\,000~km\,s$^{-1}$ are unreliable. Putting everything together, we decide to study putative B0 and B1 type HVSs with velocities between 500 and 1\,000~km\,s$^{-1}$.

Within the scenarios of interest, we set the most promissory and less common stellar spectral type (B0) for the most conservative case, that is, a star with the lowest spatial velocity (500~km\,s$^{-1}$) propagating through the Galactic disk (d). For a B0 star, plausible wind  parameters are $\dot{M}_{\rm w} = 10^{-8}~$M$_\sun$\,yr$^{-1}$ and $v_{\rm w}=$1\,500~km\,s$^{-1}$ \citep{Krticka_2014, Kobulnicky_2019}. In addition, we consider HVSs with $V_\star = 1\,000$~km\,s$^{-1}$, \mbox{$\dot{M}_{\rm w} = 10^{-9}~$M$_\sun$\,yr$^{-1}$} and $v_{\rm w}=$1\,200~km\,s$^{-1}$ propagating through three different media: the Galactic disk, the Galactic halo (h) and a molecular cloud (mc). This is motivated by the different densities in each of these media, easily reached by the HVSs, and the expectation that the FS is more luminous in a denser medium \citep{Martinez2021}. Lastly, we consider a B2 SRS with a spatial velocity of $V_\star = 60\,000$~km\,s$^{-1}$, \mbox{$\dot{M}_{\rm w} = 10^{-8}~$M$_\sun$\,yr$^{-1}$} and $v_{\rm w}=$1\,000~km\,s$^{-1}$, a viable scenario according to \cite{Dremova_2017}.

We summarise the characteristics of the selected scenarios in Table~\ref{table:systems}. Henceforth, we will refer to the systems studied as \texttt{<spectral type>-<velocity of the star [in units of $10^3$\,km\,s$^{-1}$]>-<medium of propagation>}. For instance, B1--1--d represents a B1 star propagating at 1\,000~km\,s$^{-1}$ through the Galactic disk.

\begin{table*}[h]
    \centering
    \caption[]{Parameters of the systems modelled. The values of $v_\mathrm{w}$ and $\dot{M}_{\rm w}$ are taken from \cite{Krticka_2014} and \cite{Kobulnicky_2019}, and $R_\star$ and $T_\star$ from \cite{Harmanec_1988}. We assume solar abundances.}
    \begin{tabular}{l l| c c c c c}
    \hline\hline\noalign{\smallskip}
        & & \multicolumn{5}{ c}{Scenario} \\ 
Parameter           & Symbol        & B0--0.5--d    & B1--1--h  & B1--1--d  & B1--1--mc & B2--60--d \\ 
\midrule
Peculiar velocity   & $V_\star$ [km\,s$^{-1}$]    & 500           & 1000      & 1000      & 1000      & 60\,000 \\
Ambient density     & $n_\mathrm{ISM}$ [cm$^{-3}$] & 10     & 0.1       & 10        & 100       & 10  \\
Ambient mean molecular weight & $\mu_{\rm ISM}$     & 1.28      & 1.28      & 1.28      & 2.35     & 1.28 \\
Ambient temperature [K] & T$_{\rm ISM}$    & 100       & 100       & 100       & 10       & 100 \\
Spectral type       &             & B0            & B1        & B1        & B1        & B2 \\ 
Wind velocity       & $v_\mathrm{w}$ [km\,s$^{-1}$]& 1500         & 1200      & 1200      & 1200      & 1000 \\
Wind mass-loss rate & $\dot{M}_{\rm w}$ [M$_\sun$\,yr$^{-1}$]       & $10^{-8}$     & $10^{-9}$ & $10^{-9}$ & $10^{-9}$ & $10^{-10}$ \\ 
Stellar radius      & $R_\star$ [R$_\sun$]   & 5.5           & 4.8       & 4.8       & 4.8       & 4 \\
Stellar temperature & $T_\star$ [kK]      & 29            & 25        & 25        & 25        & 20 \\
Distance to BS apex & $R_0$ [AU]          & 792         & 1120    & 112     & 26.2      & 0.54 \\
\midrule
\bottomrule
\end{tabular}
\label{table:systems}
\end{table*}


\subsection{Geometry}\label{subsec:geometria}

The BS forms as the result of the collision between the stellar wind and the ISM material. The former can be modelled as a spherical wind and the latter as a planar wind in the reference frame of the star. The FS propagates through the ISM and the RS propagates through the stellar wind, both separated by a contact discontinuity (CD). The shape of the CD is defined by the condition that the total momentum flows tangential to the shocked region, so that the flux of mass through the CD surface is null \citep{Wilkin_1996, Christie_2016}. Each system is then divided in four regions: free-flowing stellar wind, unshocked ISM, shocked ISM, and shocked stellar wind, as represented in Fig.~\ref{fig:BS}. 

The DSA mechanism operates in strong and adiabatic shocks. For massive stars, the RS always fulfils these conditions given that radiative cooling is not efficient there \citep{del_Valle_2012}. The FS, on the other hand, is radiative for typical runaway stars, but for the HVSs and SRSs considered in this work the FS is strong and adiabatic instead (see forthcoming Sec.~\ref{subsec:Hydro}). Thus, both the RS and the FS are promissory accelerators of cosmic rays and are included in the model.

The closest position of the CD to the star, known as the stagnation point, is located along the axis of symmetry where the total pressures of the ISM and the stellar wind cancel each other out. 
The thermal pressure in both the ISM and the wind is negligible. Thus, the ISM total pressure ($P_{\rm ISM}$) is:
\begin{equation}
    P_{\rm ISM} = \rho_{\rm ISM}\,V_\star^2,
    \label{Eq:P_ISM}
\end{equation}
where $\rho_{\rm ISM}$ is the density of the ISM. On the other hand, the total pressure of the stellar wind is given in terms of the mass-loss rate, the distance to the star, and the velocity of the wind:
\begin{equation}
    P_{\rm w} = \frac{\dot{M}_{\rm w} v_{\rm w}}{4\pi R(\theta)^2}.
    \label{Eq:P_w}
\end{equation}

\noindent Matching Eqs.~(\ref{Eq:P_ISM}) and (\ref{Eq:P_w}), the stagnation point is located at
\begin{equation}
    R_0 = \left[\frac{\dot{M}_{\rm w} v_{\rm w}}{4\pi\rho_{\rm ISM} V_\star^2}\right]^{1/2}.
    \label{Eq:R_0}
\end{equation}

As it will be shown in the forthcoming Sec.~\ref{sec:results}, in the case of BSs from HVSs and SRSs, NT processes are relevant even at distant regions from the apex. Thus, a one-zone model approximation in which the emitter is considered homogeneous is not appropriate if one aims at more detailed quantitative predictions. 
Therefore, we adopt a multi-zone emission model based on the one developed by \cite{del_Palacio_2018}, with the major difference being the incorporation of the FS in the model. For this, we follow an analogous approach as the one used for the RS in \cite{del_Palacio_2018} with some small modifications in the hydrodynamics. 

Each shock in the BS is treated as a two-dimensional (2D) structure. Neglecting the width of the shocked gas layers, these are 
co-spatial with the CD\footnote{Fig.~\ref{fig:BS} shows a shocked gas layer of non-zero width for illustrative purposes only.}. The shape of the CD is calculated using the formulae given by \cite{Wilkin_1996}. We assume that the shocked gas flows downstream at a fixed angle $\phi$ around the symmetry axis, dragging away NT particles. We model the 2D structure as a sum of 1D linear emitters embedded in the 3D physical space. For a given angle $\phi$ there are several 1D emitters, each starting at a different angle $\theta$ with respect to the symmetry axis where relativistic particles are injected, and consisting of multiple cells located along the path of the 1D emitter on the CD. Particles enter in the BS and are accelerated in the first cell of each 1D emitter, and then they move to the following cells up to an angle $\theta_{\rm max}$ (Fig.~\ref{fig:BS}). All the 1D emitters at a certain angle $\phi$ are summed up, thereby obtaining a 1D structure that contains all the relativistic particles along a shock at that particular angle. Finally, we rotate this 1D structure made of all the 1D emitters with the same $\phi$ value around the symmetry axis to get the full 2D structure of the BS. 
At last, we note that in all the scenarios we study, the stellar velocity is significantly below the speed of light. Therefore, relativistic effects in the hydrodynamics, as well as in the calculation of the NT particle distribution at each cell and the radiation they produce (e.g., Doppler boosting), can be neglected. 


\subsection{Hydrodynamics}\label{subsec:Hydro}

We introduce here the semi-analytical hydrodynamical model used to characterise the properties of the shocked gas in the BS. In \cite{del_Palacio_2018}, Rankine-Hugoniot jump conditions were assumed, which near the apex yields correct values up to first order. Here, we adopt a more consistent approach that is suitable to model both the RS and the FS up to distant regions from the BS apex. 
Firstly, we assume that the total pressure at $R_0$ is the ambient total pressure. Secondly, we consider Bernoulli's equation across the shock:
\begin{equation}
    \frac{1}{2}v_{\rm u}^2 + \left(\frac{\gamma_{\rm ad}}{\gamma_{\rm ad}-1}\right) \frac{P_{\rm th, u}(\theta)}{\rho_{\rm u}(\theta)} = \frac{1}{2}v(\theta)^2 + \left(\frac{\gamma_{\rm ad}}{\gamma_{\rm ad}-1}\right) \frac{P_{\rm th}(\theta)}{\rho(\theta)},
    \label{Eq:energy}
\end{equation}
where we labelled the upstream region with the subscript \textit{u}. Given that $R_0 \gg R_\star$ for the scenarios considered here (Table~\ref{table:systems}), we adopt $v_{\rm w} = v_\infty$. Then, $v_{\rm u}$ is equal to $v_\infty$ in the RS, and equal to $V_\star$ in the FS. Moreover, we neglect the thermal pressure in the upstream region when compared to the ram (kinetic) pressure, since $P_{\rm th, u} \ll P_{\rm ram, u} \propto v_{\rm u}^2$.
The incoming stellar wind and ISM impact perpendicularly to the BS apex, where the fluid halts. Hereafter, we represent quantities at the stagnation point ($\theta = 0$) with a subscript 0.
Using Eq.~(\ref{Eq:energy}), we determine the density at the stagnation point as:
\begin{equation}
    \rho_0 =
\left(\frac{\gamma_{\rm ad}}{\gamma_{\rm ad}-1}\right) \frac{2 P_{{\rm th}0}}{v_{\rm u}^2} = 5\, \frac{P_{{\rm th}0}}{v_{\rm u}^2} = 5\rho_{\rm u},
    \label{Eq:rho_0}
\end{equation}
assuming that the fluid behaves like an ideal gas with adiabatic coefficient $\gamma_{\rm ad} = 5/3$.

We assume that the shocked gas moves parallel to the CD. In the regions where the shocked fluid is subsonic, the flux of momentum that crosses the shock perpendicularly heats the downstream material, converting the total pressure in the upstream into thermal pressure in the downstream. We can thus calculate the total pressure at each point as:
\begin{equation}
     P(\theta) = P_{\rm th}(\theta) =
    \rho_{\rm u}(\theta) v_{\rm u} v_{{\rm u}\perp}(\theta).
    \label{Eq:P}
\end{equation}

\noindent On the other hand, if the fluid becomes supersonic at an angle $\theta_{\rm c}$, we consider that only the momentum density component perpendicular to the BS that crosses the BS also perpendicularly is converted to thermal energy. This two-region approach requires adopting a soft transition of the thermodynamic quantities between both regimes. We thus adopt a prescription for the total pressure for $\theta>\theta_{\rm c}$ of the form:
\begin{equation}
     P(\theta) = P_{\rm th}(\theta) =
    \rho_{\rm u}(\theta) \left(\frac{v_{\rm u}}{v_{{\rm u}\perp}(\theta_{\rm c})}\right)v_{{\rm u}\perp}(\theta)^2.
    \label{Eq:P_c}
\end{equation}

\noindent We can then calculate the density using the politropic relation:
\begin{equation}
       \rho(\theta) = \rho_0 \left(\frac{P(\theta)}{P_0}\right)^{1/\gamma_{\rm ad}}.
    \label{Eq:rho}
\end{equation}
However, at large values of $\theta$ this underestimates the fluid density, so we impose the condition $\rho(\theta) \geq \rho_{\rm min}(\theta)$, where $\rho_{\rm min}$ is defined as
\begin{equation}
    \rho_{\rm min}(\theta) =
    \begin{dcases}
    \rho_{\rm u}    \quad {\rm FS} \\
    \left(\frac{4\pi}{\Omega(\pi-\theta)} - 1\right)\rho_{\rm u}\quad {\rm RS},
    \end{dcases}
\end{equation}
with $\Omega(x) = 2\pi\left(1-\cos{(x)}\right)$, which takes into account the fraction of stellar wind accumulated in the RS up to an angle $\theta$.
To derive these expressions we have assumed that the shocked flow behaves as only one stream line which at each location adapts to the impact of the incoming upstream material, having homogeneous conditions in the direction perpendicular to the CD. This approximation is valid in the subsonic region, although it weakens in the supersonic region as the shocked flow is not causally connected along the CD. 

The assumption of a laminar flow requires the suppression of dynamical instabilities, such as Rayleigh-Taylor (triggered by density differences across the CD) and Kelvin-Helmholtz instabilities (triggered by tangential velocity differences across the CD). Such instabilities can arise in stellar BSs, especially when the ambient medium is dense \citep[e.g.][]{Comeron1998, Meyer_2016}, although a high stellar velocity inhibits their appearance up to $\theta \gtrsim 135\degree$ \citep[][]{Comeron1998, Christie_2016}. Neglecting instabilities is further justified in the presence of adiabatic shocks \citep[e.g.][]{Falceta2012}, as it is the case in the context of HVSs and SRSs. 

\begin{figure}[ht]
    \centering
    \includegraphics[angle=270, width=0.99\linewidth]{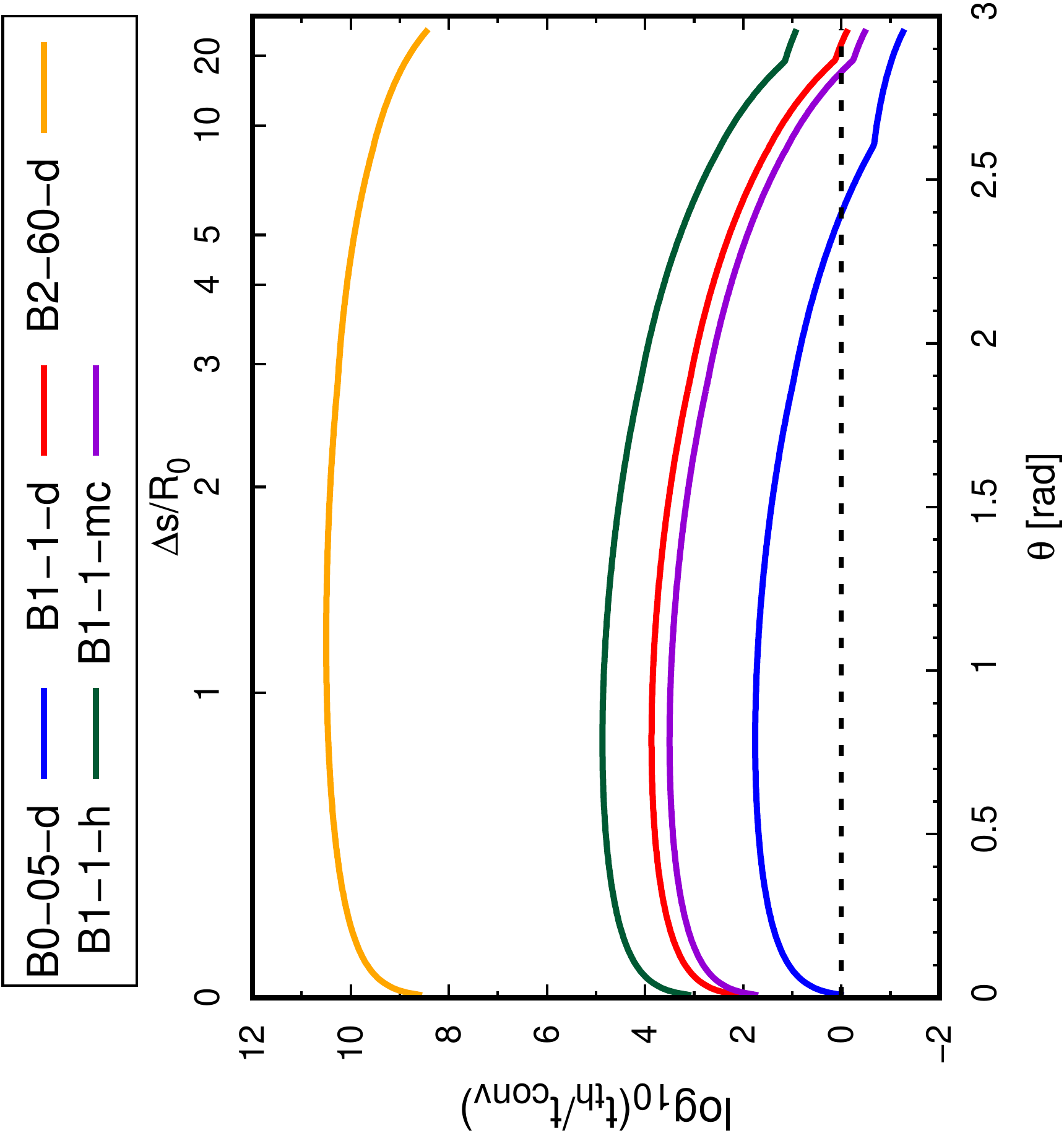}
    \caption{Logarithm of the ratio $t_{\rm th}/t_{\rm conv}$ for the different FSs studied in this work. For reference, we plot a black dotted horizontal line at zero (where $t_\mathrm{th}=t_\mathrm{conv}$); curves above this line corresponds to adiabatic shocks.}
    \label{fig:radadi}
\end{figure}

A shock is considered adiabatic when the gas escapes from the shock region before it cools significantly. This condition can be expressed as $t_{\rm th}/t_{\rm conv} \gtrsim 1$, with
\begin{equation}
    t_{\rm th}(\theta) = \frac{k_{\rm B}\,\mu_{\rm u}\,m_{\rm p}\,T}{\zeta(\theta)\,\rho_{\rm u}\,\lambda(T)},
\end{equation}
%
where the function $\lambda(T)$ depends on the temperature of the shock \citep{Myasnikov_1998}, and $m_{\rm p}$ is the proton mass. For the convection timescale we use the expression given in \cite{del_Palacio_2018}, $t_\mathrm{conv} = R(\theta)/v_{\parallel}(\theta)$. In Fig.~\ref{fig:radadi} we show the logarithm of the ratio $t_{\rm th}/t_{\rm conv}$ as a function of the angle $\theta$ for the FSs. 
This ratio increases with $\theta$ up to $\theta \sim 45\degree$ in the HVSs and up to $\theta \sim 70\degree$ in the SRS as the fluid accelerates. After this, the ratio slowly decreases as $t_{\rm conv} \propto R(\theta)$ increases. As a result, the FS is adiabatic up to $\theta \sim 160\degree$ in the systems B1--1--d and B1--1--mc, and up to $\theta \sim 135\degree$ in the system B0--05--d (see Fig.~\ref{fig:radadi}). Moreover, the FS of the systems B1--1--h and B2--60--d, as well as the RSs in all cases studied, always fulfil the adiabaticity condition. 

Lastly, the magnetic field in the subsonic regime ($\theta < \theta_\mathrm{c}$) is obtained imposing that, at each position, its pressure is a fraction $\eta_B$ of the thermal pressure of the shocked region:
\begin{equation}
    B(\theta) = \left(\eta_B \, 8\pi P_{\rm th}(\theta)\right)^{1/2}.
    \label{Eq:B_sub}
\end{equation}
In the supersonic regime ($\theta > \theta_\mathrm{c}$), we assume that the magnetic field remains frozen to the plasma, and that it is tangent to the shock surface. The later assumption is motivated by the fact that the magnetic field component perpendicular to the shock normal is amplified by adiabatic compression, and thus in general it should be the dominant component in both shocks (RS and FS).

We can then obtain the magnetic field as:
\begin{equation}
   B(\theta) = B(\theta_\mathrm{c})\, \left(
   \frac{\rho(\theta)}{\rho(\theta_\mathrm{c})}
   \frac{v(\theta_\mathrm{c})}{v(\theta)}
   \right)^{1/2}.
   \label{Eq:B_sup}
\end{equation}

\begin{figure*}[ht]
    \centering
    \includegraphics[angle=270, width=0.496\linewidth]{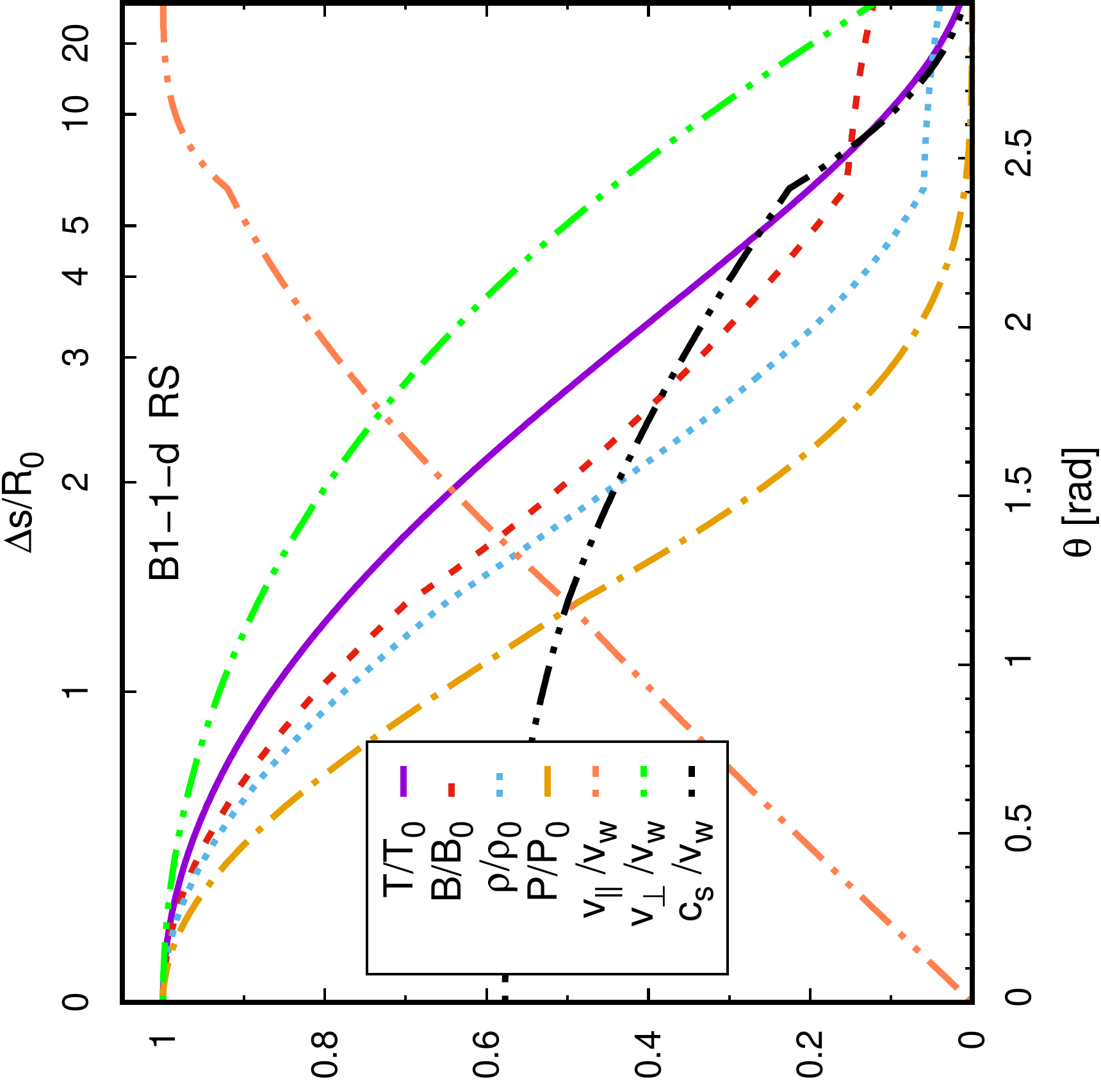}
    \includegraphics[angle=270, width=0.496\linewidth]{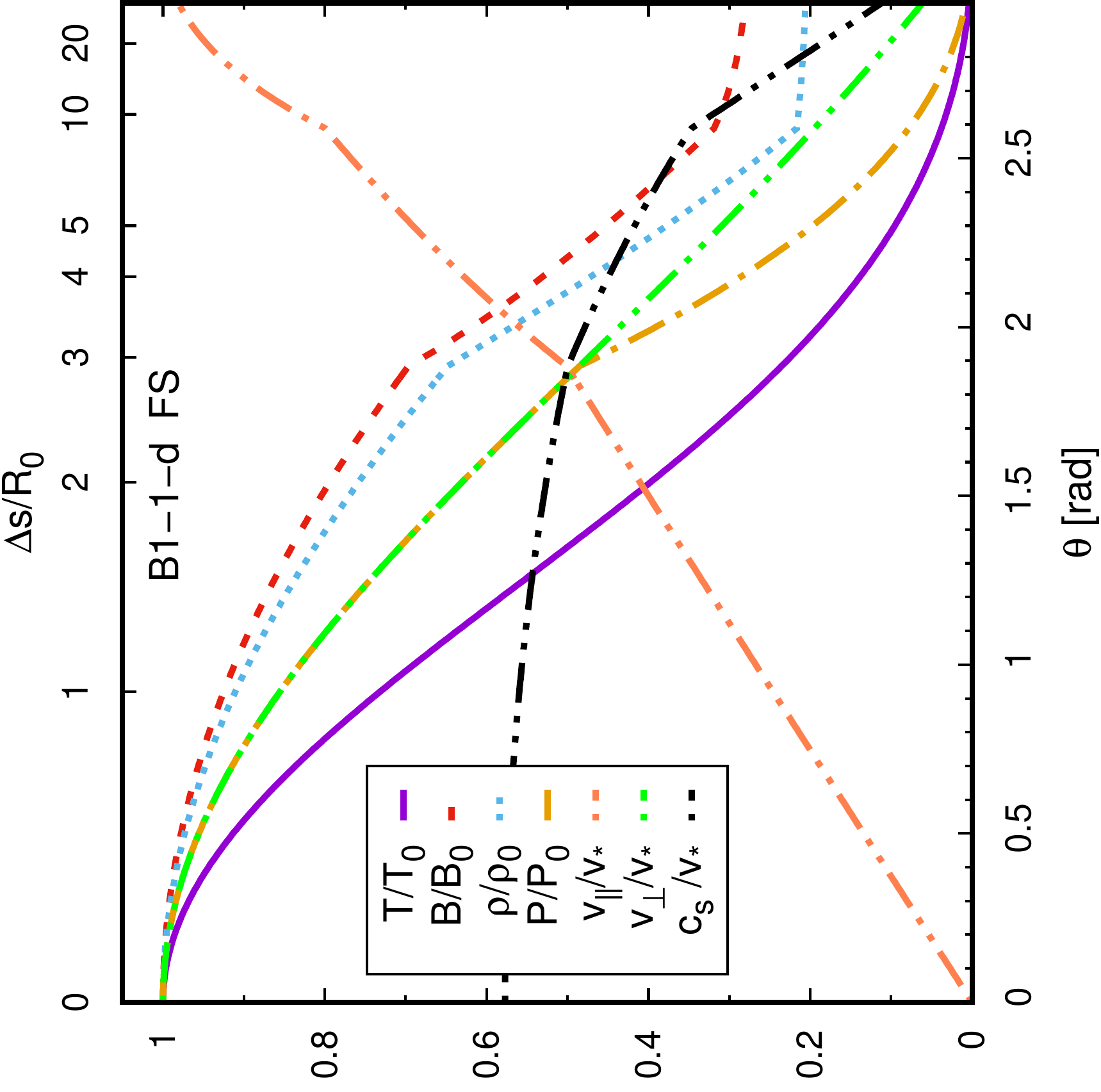}
    \caption{\emph{Left panel:} Thermodynamic variables in the RS as a function of the position angle along the shock. We also give as a reference the linear distance along the shock, $\Delta s = \int_0^\theta \mathrm{d}l$. The sub-index 0 refers to values in the apex. \emph{Right panel:} The same but for the FS.}
    \label{fig:termo}
\end{figure*}

In Fig.~\ref{fig:termo} we show the dependence of the thermodynamic quantities with $\theta$ for both shocks. 
In the apex the conversion of kinetic energy to internal energy is maximised as the shock is perpendicular. 
The tangential velocity increases monotonically with $\theta$, going from zero in the stagnation point to $v_{\rm t} \sim v_\infty$ in the RS and $v_{\rm t} \sim V_\star$ in the FS. The pressure slowly decays with $\theta$ in the FS, while it drops more abruptly in the RS because $P_{\rm RS} \propto \rho_{\rm w} \propto R^{-2}$; as a consequence, the other quantities that depend on $P$ also decay gradually. We highlight that the magnetic field decreases slowly in the FS, which favours synchrotron emission up to large values of $\theta$. Lastly, we note that this hydrodynamic model yields densities along the shocks that are slightly higher than the ones obtained assuming Rankine-Hugoniot jump conditions.
The discrepancy is a factor $\sim 1.5$ for angles $\theta \lesssim 60\degree$, and a factor $\sim 2$ in the distant regions with $\theta > 60\degree$ (see Fig.~\ref{fig:new_vs_old}). The reason for this is that, as explained above, different regions of the shocked layer affect each other making the local hydrodynamical conditions depart from the Rankine-Hugoniot ones.


\subsection{Non-thermal particles}\label{subsec:NT}

Relativistic particles can be accelerated via DSA in hypersonic and adiabatic shocks, such as the ones present in the BS (Sect.~\ref{subsec:Hydro}). Additionally, both electrons and protons could be accelerated via shock drift acceleration (SDA) in the RS, as the stellar magnetic field lines are expected to be parallel to the shock surface \citep{Marcowith2016}. Nevertheless, acquiring relativistic energies through SDA requires multiple interactions with the shock front, similarly to DSA \citep{Matthews2020}. Given that both mechanisms lead to the injection of a power-law distribution of particles and we treat the acceleration details phenomenologically, we refer in what follows only to acceleration via DSA although SDA might be also involved.

The energy distribution of the
injected particles at the \textit{i}-th cell is assumed to be
\begin{equation}
Q(E, \theta_i) = Q_0 E^{-p} \exp(-E/E_{\rm max}),
\label{Eq:Q(E)}
\end{equation}
where $Q_0$ is a normalisation factor, $p$ is the spectral index, and $E_{\rm max}$ is the cut-off energy, all dependent on $\theta_i$.

The normalisation constant $Q_0$ is set by the condition $\int E Q(E, \theta_i)\,{\rm d}E = \Delta L_{\rm NT}(\theta_i)$, being $\Delta L_{\rm NT}(\theta_i)$ the power available to accelerate NT particles at each position. That is,
\begin{equation}
    \Delta L_{\rm NT}(\theta_i) = f_{\rm NT}\,\Delta L_{\perp}(\theta_i) = f_{\rm NT}\, S_{\rm E}(\theta_i)\, A_{\perp}(\theta_i),
\label{Eq:L_NT}
\end{equation}
where $S_{\rm E}(\theta_i)$ is the energy flux per unit volume of the corresponding fluid, and $A_{\perp}(\theta_i)$ is the area of the cell surface projected perpendicular to $\vec{v}_{\rm u}$. This area is calculated as \mbox{$A_{\perp}(\theta_i)= R(\theta_i)\,\sin(\theta_i)\,\Delta l(\theta_i)\,\sin{(\alpha_i)}\,\Delta \phi$}, with $\Delta l(\theta_i)$ being the length of the cell and  $\alpha_i$ the angle between $\vec{v_{\rm u}}$ and the tangent to the shock, $\vec{T}$ (Fig.~\ref{fig:BS}). The parameter $f_{\rm NT}$ is defined as the fraction of the power injected to the BS that goes to NT particles. We adopt $f_{\rm NT} = 0.1$ and assume that 95\% of this power goes to protons, while the remaining 5\% goes to electrons. Finally, the energy flux per unit volume in the subsonic regime is $S_{\rm E} = 0.5\,\rho_{\rm u}\,v_{\rm u}^3$, whereas in the supersonic regime only the perpendicular velocity component is converted into thermal energy and therefore $S_{\rm E} = 0.5\,\rho_{\rm u}\,v_{\rm u}^2\,v_{\rm u \perp}$. Contrary to what happens in the RS, the injection of energy in the FS is relevant even at large values of $\theta$, where the flow is more susceptible to develop instabilities. That could potentially increase the area of the shock, thus enhancing the injected power and, consequently, the emitted luminosity \citep[e.g.][]{delaCita2017}. Nevertheless, the variability induced by this effect is not expected to dominate the average luminosities predicted in our model.

We can determine $p$ in terms of the compression factor $\zeta$ as \citep[e.g.][]{Caprioli2014}:
\begin{equation}
    p(\theta_i) = \frac{\zeta(\theta_i) + 2}{\zeta(\theta_i)-1} , \quad \zeta(\theta_i) = \left(\frac{\gamma_{\rm ad}+1}{\gamma_{\rm ad}}\right)\frac{M(\theta_i)^2}{M(\theta_i)^2+2},
    \label{Eq:indice_espectral}
\end{equation}
where $M$ is the Mach number. Finally, $E_{\rm max}$ is obtained by equating the acceleration and energy loss timescales.

The steady-state particle distribution at the injection cell is:
\begin{equation}
    N_0(E, \theta_i) \approx Q(E, \theta_i) \times {\rm min}(t_{\rm cell}, t_{\rm cool}),
    \label{Eq:N_0}
\end{equation}
being $t_{\rm cell}$ the cell convection time and $t_{\rm cool}$ the cooling timescale. In the BS, electrons cool mainly by synchrotron and inverse Compton (IC) interactions, the latter with both the stellar UV (IC$_\star$) and dust IR (IC$_{\rm IR}$) radiation fields \citep{Martinez2021}. Adiabatic losses can also be relevant, whereas relativistic Bremsstrahlung losses are negligible. Protons cool by proton-proton inelastic collisions, a rather minor effect, and adiabatic losses. 

The NT particles are confined within the shock and so they are dragged by the fluid. This occurs because the particles gyro-radii, $r_{\rm g}(E, \theta) \propto E/B(\theta)$, even for $E \sim E_\mathrm{max}$ are much smaller than the shocked layer width (the shock typical scale) $H(\theta)$. The latter is calculated considering mass conservation across the shock
\begin{equation}
   H(\theta) = \frac{\int_0^\theta \rho_{\rm u}(\theta')\,v_{\rm u}\,A_\perp(\theta')\,{\rm d}\theta'}{2\pi\,R(\theta)\,\sin{(\theta)}\,\rho(\theta)\,v_\parallel(\theta)},
   \label{Eq:H}
\end{equation}
yielding a typical value of $H(\theta) \sim 0.3 R(\theta)$ for $\theta < \pi/2$ \citep{Christie_2016}. Under these conditions, let us consider that there is a certain number of NT particles with energy $E$ in the $i$-th cell. By the time they reach the ($i$+1)-th cell, their energy will be $E' \leq E$, and the size and convection velocity of the cell will also be different. Nonetheless, in the steady-state, the flux of particles in the position and energy space must be conserved. Considering this, we obtain the evolution of the particle energy distribution along a linear emitter\footnote{We note that the inclusion of  $t_\mathrm{cell}$ in this expression is a small correction to the one used in \cite{del_Palacio_2018}.}:
\begin{equation}
        N(E', i+1) = N(E, i) \, \frac{ |\dot{E}(E,i+1)|}{ |\dot{E}(E',i+1)|} \,  \frac{t_\mathrm{cell}(i+1)}{t_\mathrm{cell}(i)},
       \label{Eq:N(E)}
\end{equation}
with $\left|\dot{E}(E, i)\right| = E/t_{\rm cool}(E, i)$ the cooling rate for particles of energy $E$ at the position $\theta_i$ and $t_{\rm cell}(i)$ the convection time of the \textit{i}-th cell. Finally, the energy $E'$ is given by the condition $t_{\rm cell} = \int_E^{E'} \dot{E}(\tilde{E}, i)\,{\rm d}\tilde{E}$. 

We use the formulae given by \cite{Khangulyan_2014} to calculate the isotropic IC cooling timescales, given that the electron distribution is isotropic at each position and the fluid is non-relativistic. These expressions take into account the Klein-Nishina (K-N) cross section for the interaction at high energies. For the case of the stellar radiation field, we consider the star as a black body emitter with temperature $T_\star$ and a dilution factor of the photon field $\kappa_\star = \left[R_\star/(2R(\theta))\right]^2$. For the case of the IR photon field produced by the dust, we assume isotropy within the NT emitter, and we approximate its spectrum with a Planck law of temperature $T_{\rm IR} = 100$~K. The corresponding dilution factor is $\kappa_{\rm IR}(\theta) = U_{\rm IR}(\theta)/U_{\rm BB}$, where $U_{\rm BB}$ is the energy density of the radiation of a black body with temperature $T_\mathrm{IR}$. Since $U_{\rm IR} \approx L_{\rm IR}/\left(4\pi R(\theta)^2 c\right)$ (considering that the dust is concentrated in a thin shell surrounding the BS) and $U_{\rm BB} = 4 \sigma T_{\rm IR}^4/c$, we obtain:
\begin{equation}
    \kappa_{\rm IR} = \frac{L_{\rm IR}}{16\pi \sigma T_{\rm IR}^4 R(\theta)^2}.
    \label{Eq:kappa_IR}
\end{equation}

\section{Results} \label{sec:results}

First, we compute the timescales and particle energy distributions for the scenarios studied. Then, we use a one-zone model to estimate the luminosity scaling with the relevant parameters of the systems. Finally, we present the spectral energy distributions (SEDs), and discuss the detectability of each system.

The magnetic field in the shocked region can be generated by adiabatic compression of the ISM (star) magnetic field in the FS (RS) and/or be amplified by the action of cosmic rays. Adopting $\eta_{\rm B} = 0.1$ in Eq.~(\ref{Eq:B_sub}) yields values of $B$ consistent with a ratio between NT energy density and magnetic energy density of $U_{\rm NT}/U_{\rm B} \gtrsim 1$. Thus, in both shocks the magnetic field could be the result of amplification by cosmic rays \citep{Bell2004}. Alternatively, in the RS the magnetic field could come from adiabatic compression of the stellar magnetic field. Under these conditions, and adopting an Alfv\'en radius $r_{\rm A} \sim R_\star$, the stellar magnetic field in the stellar surface is $B_\star \sim 0.25\,B(\theta)\,\left(R(\theta)/R_\star \right)\,\left(v_\infty/v_{\rm rot}\right)$, with the star rotation speed being $v_{\rm rot} \sim 0.1 \,v_\infty$ \citep{del_Palacio_2018}. We obtain $B_\star \sim 25$~G for the B0 star, $B_\star \sim 10$~G for the B1 stars, and $B_\star \sim 3$~G for the B2 star. Given that these are plausible values \citep{Parkin2014}, adopting $\eta_{\rm B} = 0.1$ is a reasonable assumption. Additionally, we consider an equipartition scenario ($\eta_B = 1$) to set an upper limit to the predicted radio fluxes. At last, given that the majority of power is injected within $\theta < 170\degree$, and that instabilities can be significant for large values of $\theta$, we fix $\theta_{\rm max} = 170\degree$.

\begin{table}[ht]
\centering
\caption[]{Power injected in NT particles (electrons and protons) in each system, assuming $f_\mathrm{NT}=0.1$ and $\eta_{\rm B} = 1$ (see text in Sec.~\ref{subsec:NT} for details).}
\begin{tabular}{c| c c c}
\hline\hline       
     & \multicolumn{3}{c}{$L_{\rm NT}$ [erg\,s$^{-1}$]} \\ \cline{2-4}
Scenario      & RS    & FS       & Total     \\
\hline
B0--05--d     & $5.3 \times 10^{32}$  & $6.9 \times 10^{32}$    & $1.2 \times 10^{33}$  \\          
B1--1--h      & $3.4 \times 10^{31}$  & $1.1 \times 10^{32}$    & $1.4 \times 10^{32}$ \\         
B1--1--d      & $3.4 \times 10^{31}$  & $1.1 \times 10^{32}$    & $1.4 \times 10^{32}$ \\            
B1--1--mc     & $3.4 \times 10^{31}$  & $1.1 \times 10^{32}$    & $1.4 \times 10^{32}$ \\            
B2--60--d     & $2.3 \times 10^{30}$  & $5.5 \times 10^{32}$    & $5.5 \times 10^{32}$ \\        
\hline
\end{tabular}
\label{Tabla:L_NT}
\end{table}
\subsection{Relativistic particle population}
 In Table~\ref{Tabla:L_NT} we show the power injected in NT particles (both electrons and protons) for each scenario. This quantity increases with younger spectral types and the velocity of the star, as expected from Eq.~(\ref{Eq:L_NT}). 
 Considering that there are at most tens of thousands of HVSs in the Milky Way \citep{Marchetti_2019}, the total luminosity injected in NT particles by these sources in the Galaxy is likely below $10^{38}$~erg\,s$^{-1}$. Then, HVSs do not significantly contribute to the Galactic cosmic ray population, which has a much larger (by a factor $\sim 10^3$) contribution from supernova remnants. 

\subsubsection{Forward shock}

\begin{figure}
    \centering
    \includegraphics[angle=270, width=0.49\textwidth]{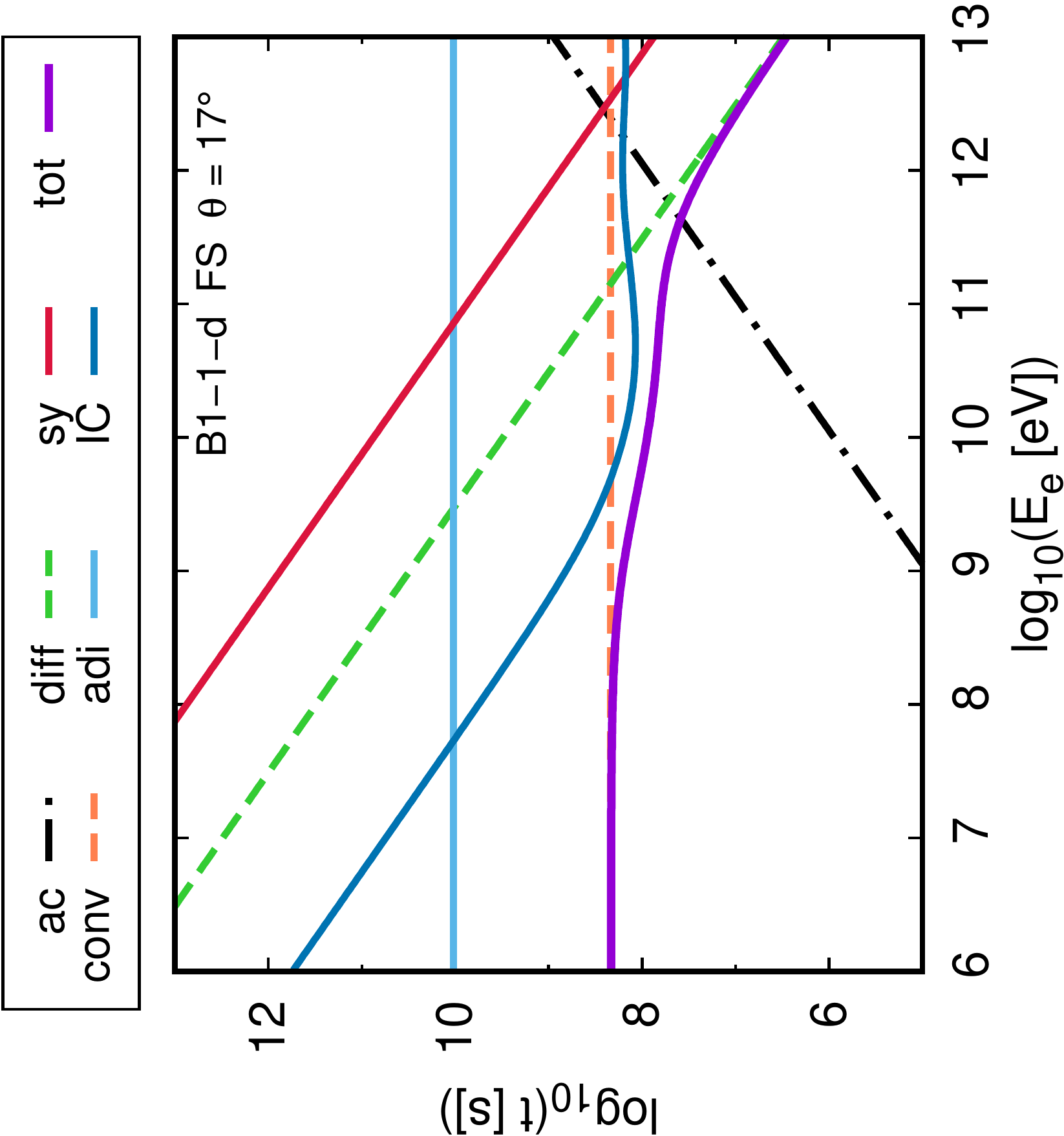}\\
    \includegraphics[angle=270, width=0.49\textwidth]{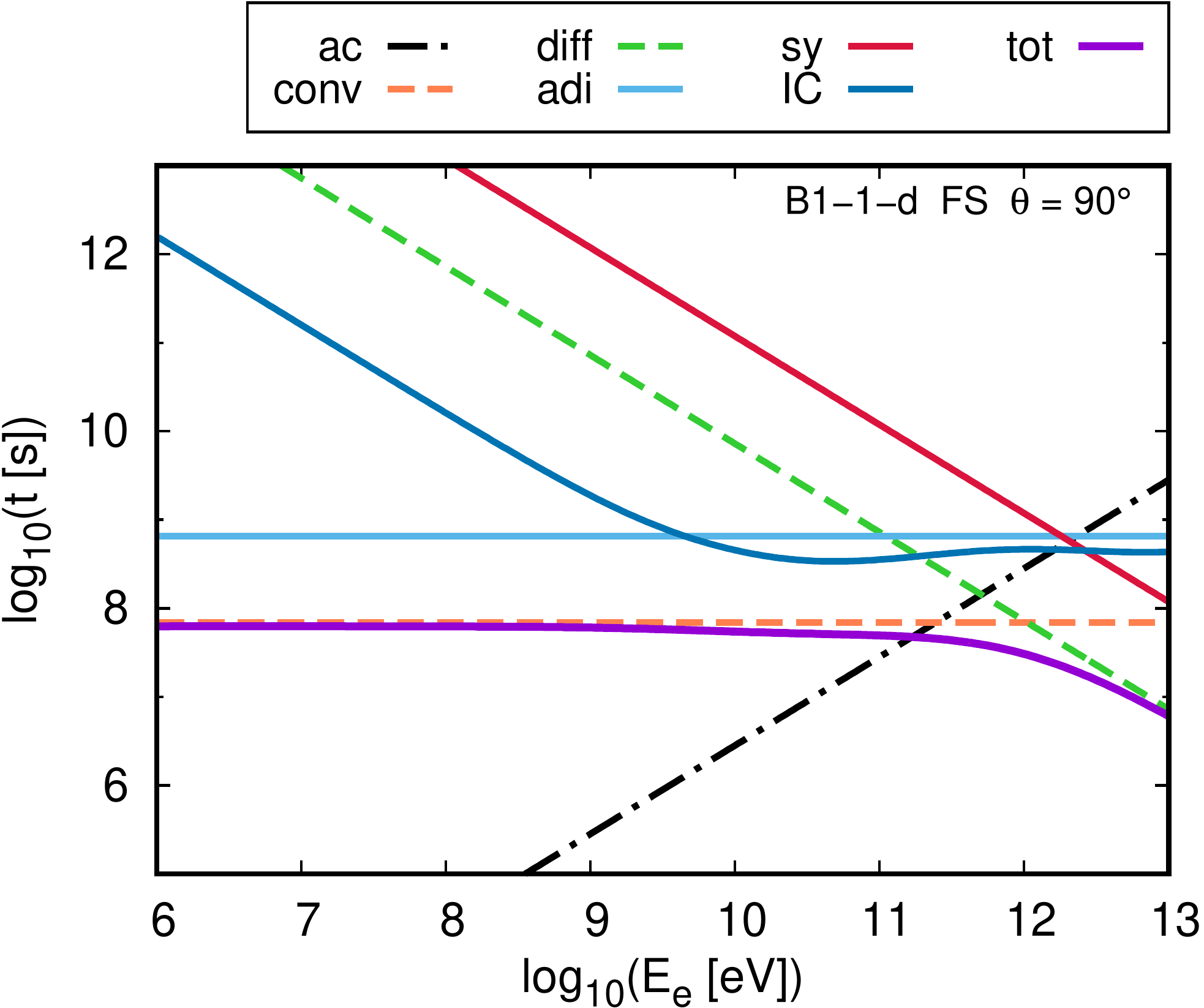}\\
    \includegraphics[angle=270, width=0.49\textwidth]{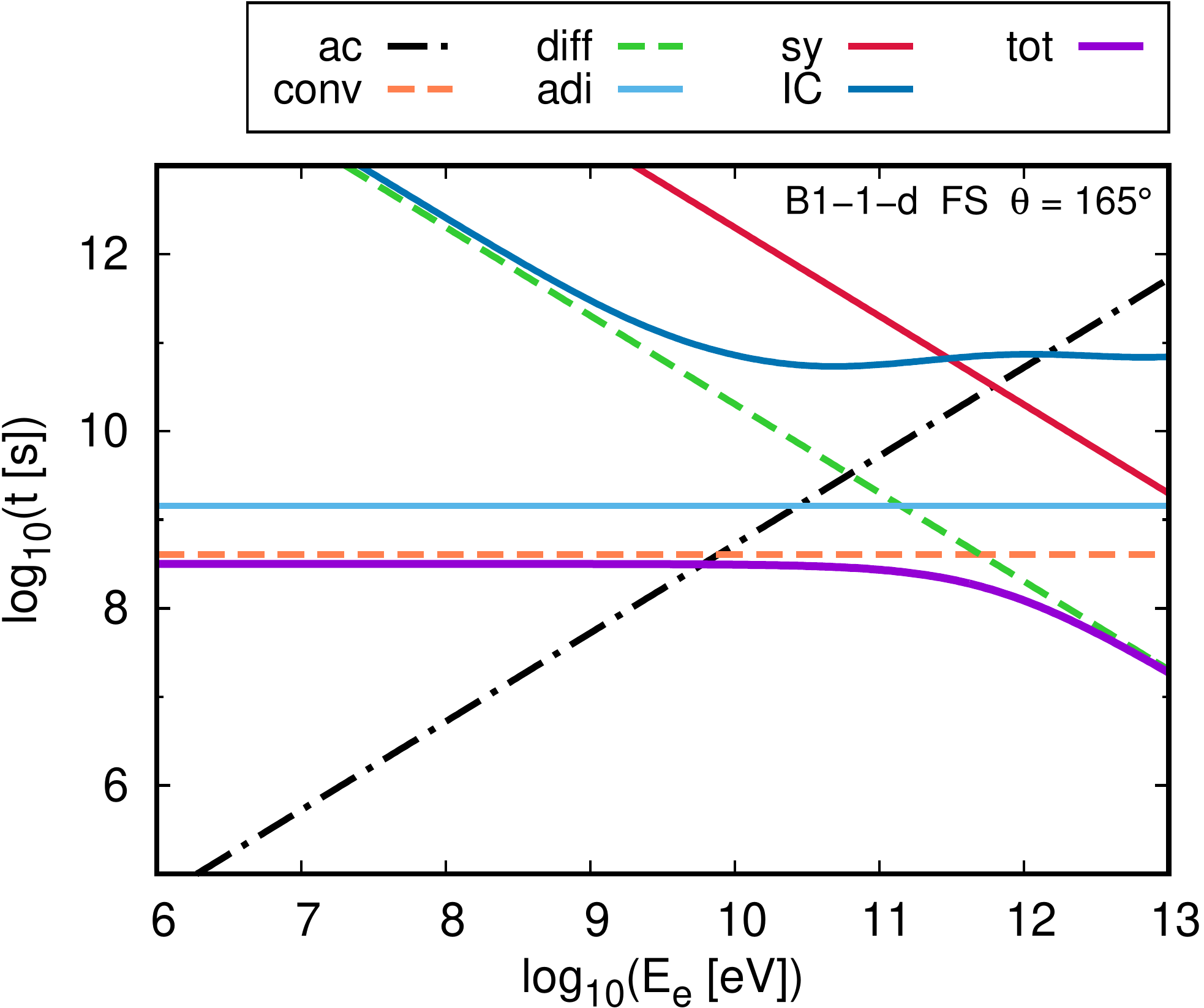}
    \\
    \caption[]{Cooling times for three different positions of electrons for the FS of the system B1-1-d. The IC cooling timescale of electrons takes into account both the stellar and dust photon fields.}
    \label{fig:tiempos_e}
\end{figure}

\begin{figure}[t]
    \centering
    \includegraphics[angle=270, width=0.49\textwidth]{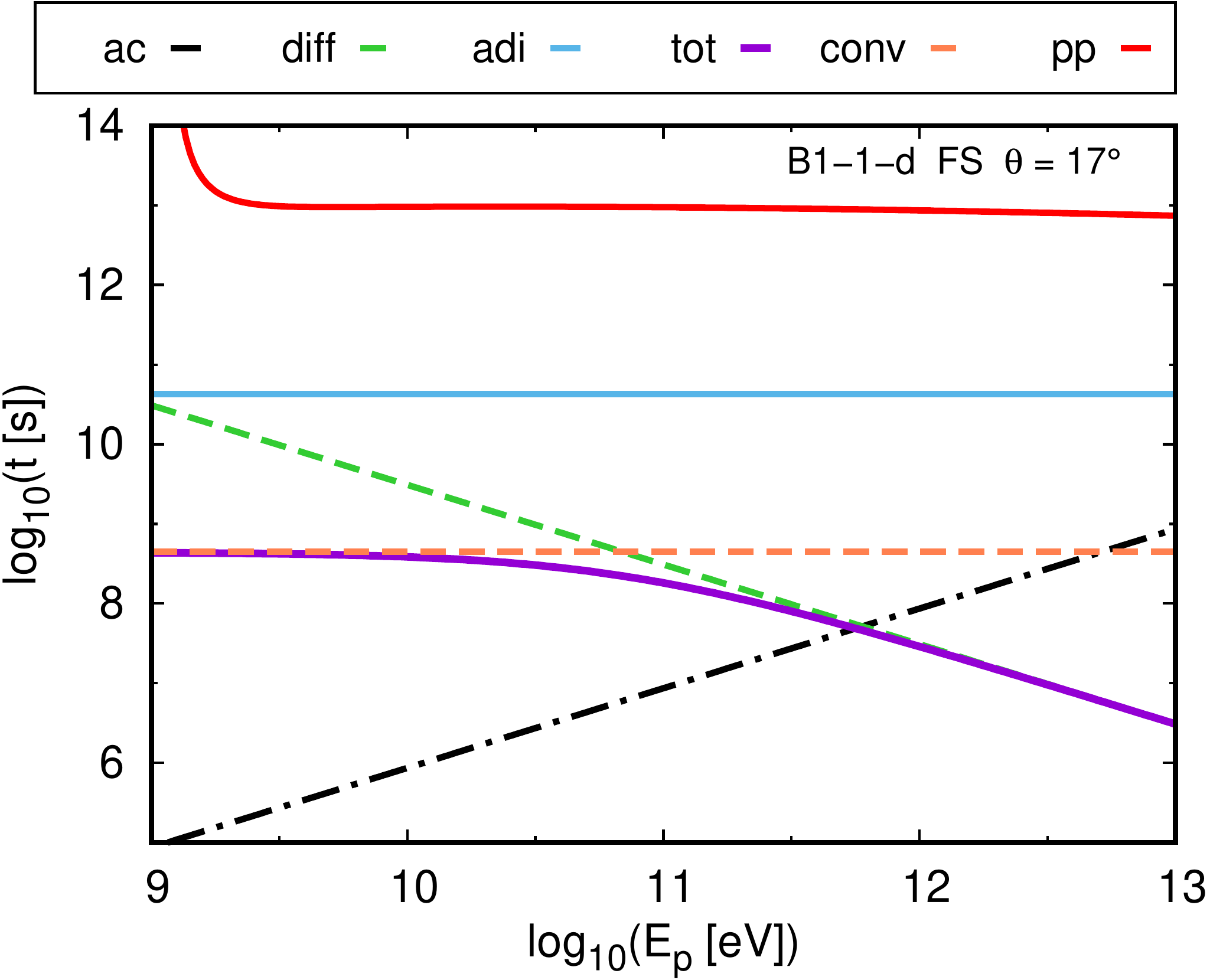}
    \caption[]{Cooling times for protons of the system B1--1--d near the apex.}
    \label{fig:tiempos_p}
\end{figure}

An example of the timescales considered is shown in Figs.~\ref{fig:tiempos_e} and \ref{fig:tiempos_p} for the FS of the system B1--1--d. For electrons near the apex, convection (escape) dominates up to $E_{\rm e} \leq 4$~GeV. In the range $4~{\rm GeV} \leq E_{\rm e} \lesssim 160$~GeV,  IC losses with the stellar photon field are dominant, also versus IR IC losses; these interactions occur in the Thomson regime for
$E_{\rm e} \lesssim 30$~GeV, and in the K-N regime for
$E_{\rm e} \gtrsim 30$~GeV. At last, diffusion (escape) losses dominate above 160~GeV, and electrons reach energies of $E_{\rm e,max} \sim 500$~GeV. In a scenario with a lower ISM density (system B1--1--h), the stagnation point is further from the star, and so the density of stellar photons in the BS is smaller. As a consequence, IC timescales are larger, and convection losses are dominant up to $E_{\rm e} \lesssim 100$~GeV, and diffusion losses are dominant in the range $100~{\rm GeV} \lesssim E_{\rm e} \lesssim 300~{\rm GeV}\sim E_{\rm e,max}$. Similarly, $R_0$ also increases for main-sequence stars with younger spectral types, as $\dot{M}_{\rm w}$ and $v_{\rm w}$ increase. Finally, for a SRS (B2--60--d), $R_0$ is significantly closer to the star ($R_0 \approx 30~R_\star$). Then, IC$_\star$ interactions are dominant for $3~{\rm GeV} \leq E_{\rm e} \lesssim 600~{\rm GeV}$, and IC$_{\rm IR}$ interactions are dominant in the range $600~{\rm GeV} \lesssim E_{\rm e} \lesssim 2~{\rm TeV}$. Above $E_{\rm e} \geq 2~{\rm TeV}$ and up to $E_{\rm max}$ diffusion dominates. Since $B(\theta) \propto P(\theta) \propto V_\star^2 \gg 1\,000$~km\,s$^{-1}$, the acceleration of NT particles is very efficient, yielding maximum energies of $E_{\rm e,max} \sim 5$~TeV for electrons and $E_{\rm p,max} \sim 10$~TeV for protons. 

In distant regions from the apex, the stellar photon field is more diluted and therefore IC losses are less important. Additionally, the convection timescale shortens as $v_\parallel$ increases. As a consequence, convection losses are completely dominant in the FS of all systems for angles $\theta \gtrsim 100\degree$. Moreover, particle acceleration is less efficient in distant regions since both $v_\perp$ and $B$ decrease with $\theta$ (Fig.~\ref{fig:termo}). Then, $E_{\rm max}$ diminishes with $\theta$ for both electrons and protons.

In contrast, protons are convected away from the FS without radiating a significant fraction of their energy. For the HVSs, close to the apex convection losses are the dominant process for protons with $E_{\rm p} < 160$~GeV. In the case of the SRS (scenario B2--60--d), convection dominates up to $E_{\rm p} \gtrsim 1$~TeV\footnote{ We note that after convecting away from the BS, protons diffuse in the surrounding medium and, in dense environments, can produce significant $\gamma$-ray radiation via proton-proton collisions \citep{delValle2014}.}. Above the energies mentioned, and up to $E_{\rm p,max}$, diffusion losses are dominant.

Protons reach their maximum energy at the apex ($\theta=0$). We estimate the scaling of $E_{\rm p,max,0}$ with the system parameters by matching $t_{\rm ac,0}(E_{\rm p}) = t_{\rm diff,0}(E_{\rm p})$. Considering diffusion in the Bohm regime, these timescales are:
\begin{align}
    t_{\rm ac,0}(E_{\rm p}) &= \frac{2\pi c}{q}\frac{E_{\rm p}}{B_0\,v_{{\perp}0}} = \frac{2\pi c}{q}\frac{E_{\rm p}}{B_0\,v_{\rm u}},\label{Eq:t_ac}
    \\
    t_{\rm diff,0}(E_{\rm p}) &= \frac{R_0^2}{2D_{\rm B,0}} = \frac{3q}{2c}\frac{R_0^2\,B_0}{E_{\rm p}},\label{Eq:t_diff}
\end{align}

\noindent where $c$, $q$ and $D_{\rm B}$ are the speed of light, proton charge and diffusion coefficient, respectively. Finally, using Eqs.~(\ref{Eq:P}) and (\ref{Eq:B_sub}), we get $E_{\rm max,p,0} \propto R_0\,B_0\,V_\star^{0.5} \propto \dot{M}_{\rm w}^{0.5}\,v_{\rm w}^{0.5}\,V_\star^{0.5}$. This maximum energy is $E_{\rm max,p} \sim 1$~TeV for the HVSs and $E_{\rm max,p} \sim 10$~TeV for the SRS.

\begin{figure*}
    \centering
    \includegraphics[angle=270, width=0.49\textwidth]{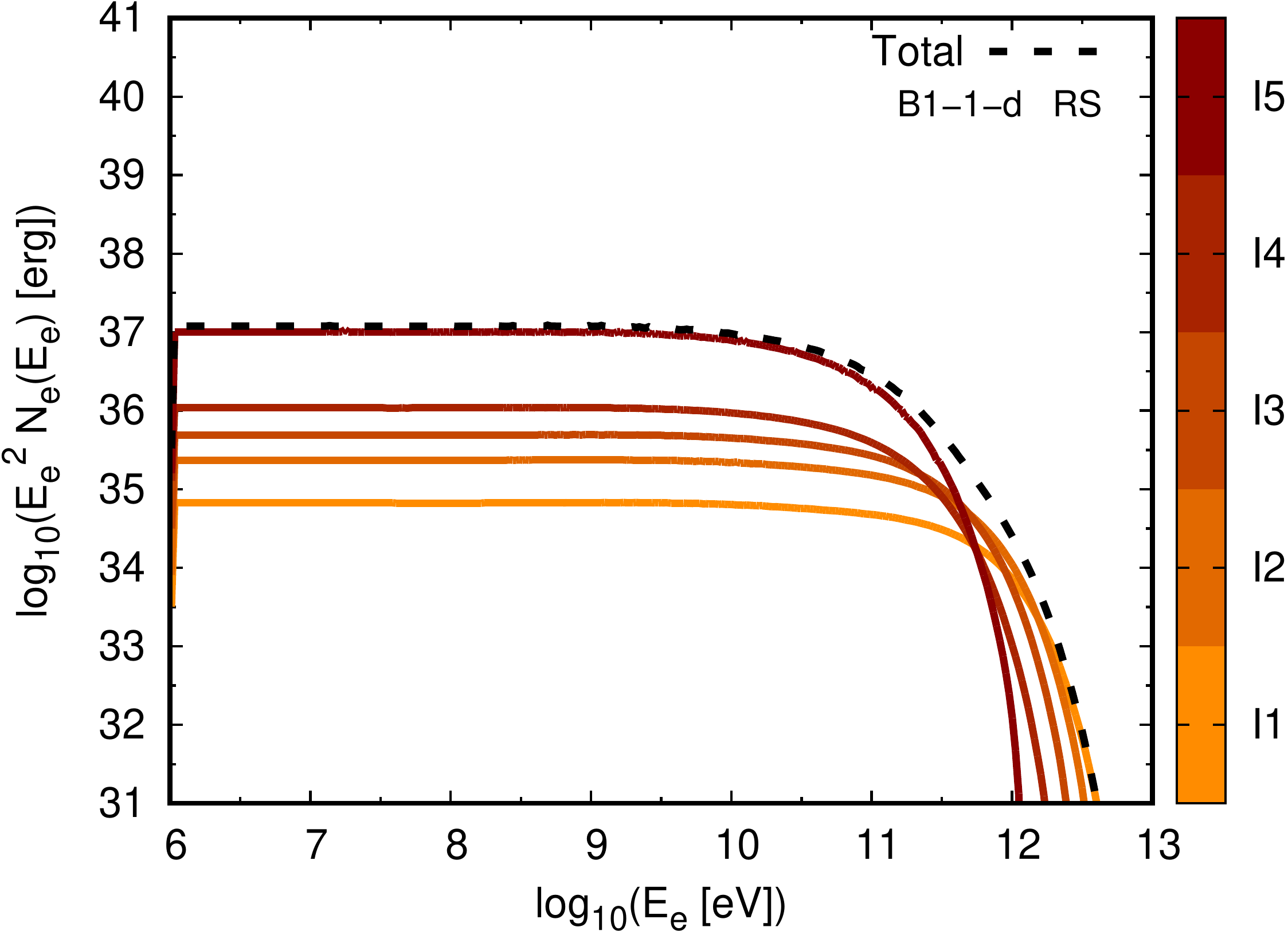}
    \includegraphics[angle=270, width=0.49\textwidth]{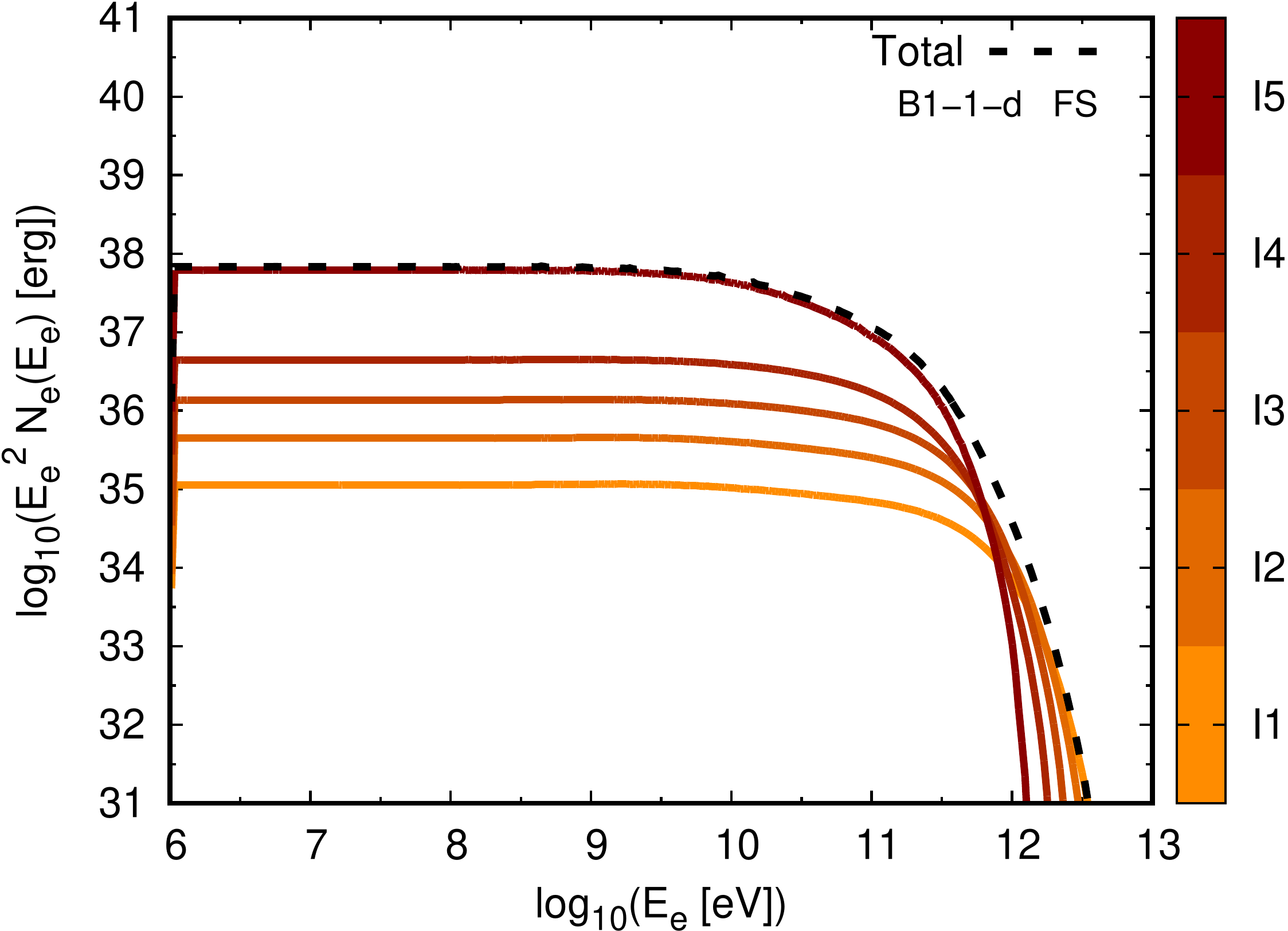}\\
    \includegraphics[angle=270, width=0.49\textwidth]{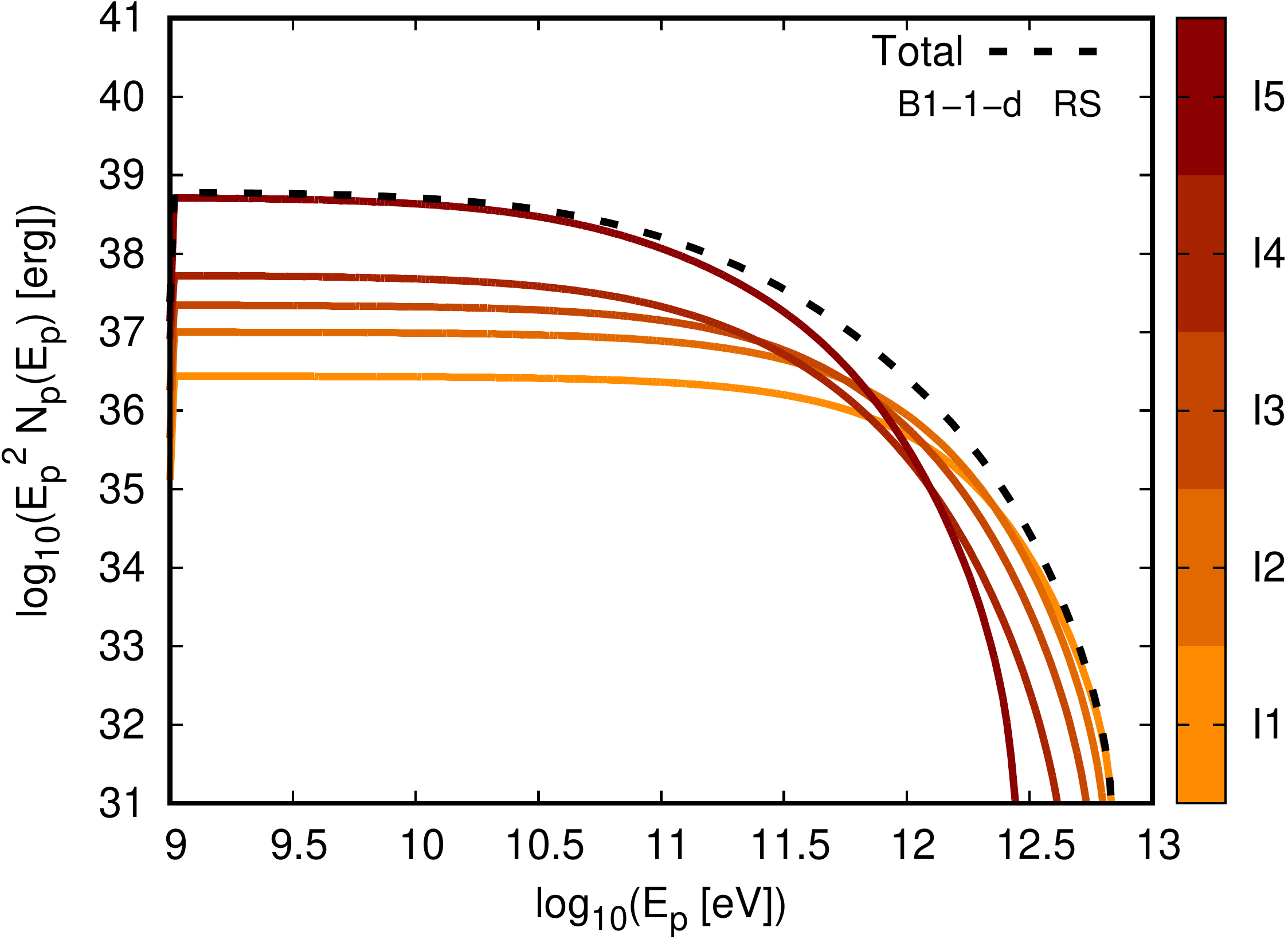}
    \includegraphics[angle=270, width=0.49\textwidth]{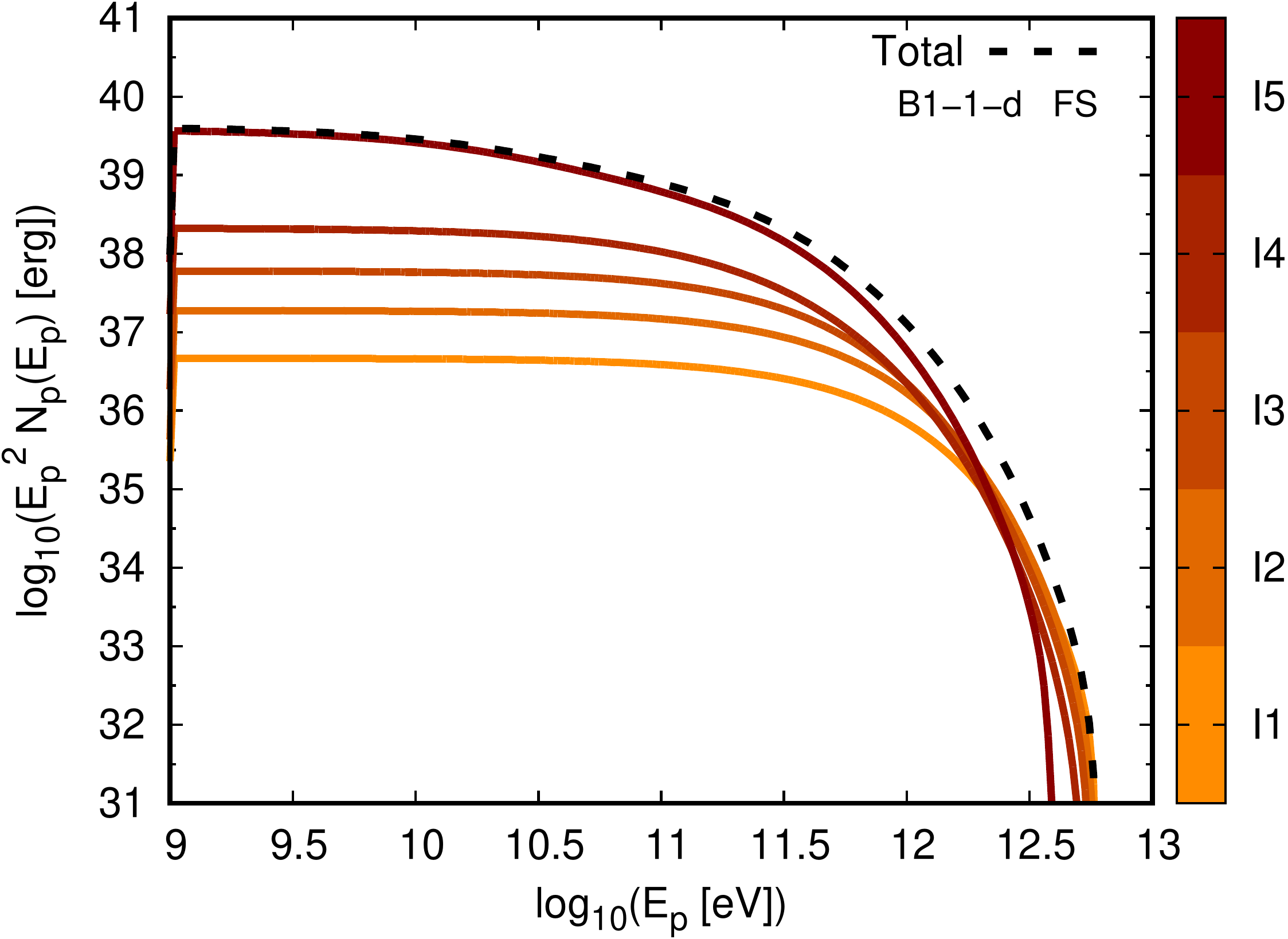}\\
    \caption[]{Particle energy distribution for electrons (top panels) and protons (bottom panels). The left panels are for the RS and the right panels are for the FS. The colour scale represents five different regions of the emitter that correspond to intervals of length $\Delta \theta = \theta_{\rm max}/5 = 34\degree$. The black dashed line corresponds to the total particle energy distribution. }
    \label{fig:distribuciones}
\end{figure*}

When convection dominates, particles move along the BS with an energy distribution that keeps the same spectral index as the injected distribution.
On the other hand, when IC in the Thomson regime or diffusion dominates, the particle energy distribution is softened. In Fig.~\ref{fig:distribuciones} we show the particle energy distribution of electrons and protons in different regions of the RS and the FS for the system B1--1--d.

\subsubsection{Reverse shock}

Characteristic timescales near the apex for the RS are similar to the ones of the FS of the HVSs. Nonetheless, the behaviour is different for the SRS. Despite the acceleration timescale decreases as the magnetic field is stronger, the IC cooling timescale decreases more drastically, yielding maximum electron energies of $E_{\rm e,max} \sim 10$~GeV in the RS (the RS shock velocity is much lower than in the FS). Moreover, we highlight that IC cooling is still relevant up to $\theta \sim 160\degree$ for the RS of this system.

Finally, matching Eq.~(\ref{Eq:t_ac}) and (\ref{Eq:t_diff}) for the RS we find for protons that $E_{\rm max,p}(0) \propto R_0\,B_0\,V_{\rm w}^{0.5} \propto \dot{M}_{\rm w}^{0.5}\,v_{\rm w}$. In this case, the maximum energy is $E_{\rm max,p,0} \gtrsim 1$~TeV for the HVSs and $E_{\rm max,p,0} \sim 200$~GeV for the SRS.

\subsection{Analytical estimates on emissivity scaling}\label{subsec:estimaciones}

A one-zone approximation is suitable to obtain order-of-magnitude estimates of the radiative outputs, with the advantage that the dependencies of the emission with respect to different system parameters become explicit \citep[e.g.][]{del_Palacio_2018}. We therefore apply this formalism to derive the scaling of the BS luminosity with the relevant system parameters.  

We consider that the BS has an effective surface of size $R_\mathrm{eff} = a\, R_0$. This is mostly relevant for the FS, for which $a\sim5$ (as explained below), whereas for the RS it is $a \sim 1$. If $t_{\rm rad}$ is the cooling timescale of the dominant NT mechanism in a certain energy range, we estimate the NT power emitted in that range by each shock as
\begin{equation}
    L_{\rm rad} \sim L_{\rm NT} \,\left(\frac{t_{\rm conv}}{t_{\rm rad}}\right) \propto
    \begin{dcases}
    f(a)\,\dot{M}_{\rm w}\,v_{\rm w}^2\,\left(\frac{t_{\rm conv}}{t_{\rm rad}}\right) &{\rm RS} \\
    \rho_{\rm ISM}\,R_{\rm eff}^2\, V_\star^3\left(\frac{t_{\rm conv}}{t_{\rm rad}}\right) &{\rm FS},
    \end{dcases}
    \label{Eq:L_mec}
\end{equation}

\noindent where $f(a)$ is an order unity function that tends to $f(a) \approx 0.6$ as $a$ increases. Qualitatively, the ratio $t_{\rm conv}/t_{\rm syn}$ determines the NT luminosity emitted in the radio band. On the other hand, $\gamma$-ray emission depends on the ratio $t_{\rm conv}/t_{\rm IC\star}$, as IC process with the stellar radiation field is the dominant NT process for those energies. We mote that these timescale-ratio dependencies of the radiation luminosity are strictly valid for a dominant $t_{\rm conv}$.

The convection (escape) timescale is roughly determined by the effective radius and the sound speed in the shocked gas:
\begin{equation}
    t_{\rm conv} \sim \frac{R_{\rm eff}}{c_{\rm s}} \propto
    \begin{dcases}
    a\, \dot{M}_{\rm w}^{0.5}\,v_{\rm w}^{-0.5}\,n_{\rm ISM}^{-0.5}\,\mu_{\rm ISM}^{-0.5}\,V_\star^{-1}   &{\rm RS}\\ \\
    a\, \dot{M}_{\rm w}^{0.5} \, v_{\rm w}^{0.5}\,n_{\rm ISM}^{-0.5}\,\mu_{\rm ISM}^{-0.5}\,V_\star^{-2}   &{\rm FS}.
    \end{dcases}
    \label{Eq:t_conv}
\end{equation}

\noindent On the other hand, the synchrotron cooling timescale is \citep[e.g.][]{Blumenthal1970}
\begin{equation}
   t_{\rm syn} \propto B^{-2} \propto
   \begin{dcases}
   \left(\dot{M}_{\rm w}\,v_{\rm w}\,R_{\rm eff}^{-2}\right)^{-1} \propto a^2\,n_{\rm ISM}^{-1}\,\mu_{\rm ISM}^{-1}\,V_\star^{-2}  &{\rm RS} \\
    \left(\rho_{\rm ISM}\,V_\star^2\right)^{-1} \propto n_{\rm ISM}^{-1}\,\mu_{\rm ISM}^{-1}\,V_\star^{-2} &{\rm FS}.
   \end{dcases}
   \label{Eq:t_syn}
\end{equation}

\noindent Thus, from Eqs.~(\ref{Eq:L_mec})--(\ref{Eq:t_syn}) we obtain
\begin{equation}
    L_{\rm syn} \sim L_{\rm NT} \,\left(\frac{t_{\rm conv}}{t_{\rm syn}}\right) \propto
    \begin{dcases}
    f(a)\,a^{-1}\,\dot{M}_{\rm w}^{1.5}\,v_{\rm w}^{1.5}\,n_{\rm ISM}^{0.5}\,\mu_{\rm ISM}^{0.5}\,V_\star     &{\rm RS} \\ \\
    a^3\,\dot{M}_{\rm w}^{1.5}\,v_{\rm w}^{1.5}\,n_{\rm ISM}^{0.5}\,\mu_{\rm ISM}^{0.5}\,V_\star &{\rm FS}.
    \end{dcases}
\end{equation}

\noindent Therefore, synchrotron emission depends mostly on the stellar parameters: massive stars are promissory radio emitters, as they have fast, powerful winds. To a lesser extent, synchrotron luminosity increases with the speed of the star and the medium density. Finally, we note that the FS luminosity has a stronger dependency on the effective radius of the BS than the RS luminosity. This supports the need to incorporate the factor $a$ when using one-zone models to estimate the luminosity of BSs in systems for which the FS contribution can be significant or even dominant. A value of $a\sim5$ yields luminosities that match within a factor two or three with those obtained using a more precise multi-zone model.

The IC$_\star$ cooling timescale depends on the energy density of the stellar photon field: $t_{{\rm IC}\star}^{-1} \propto U_\star \propto L_\star R_{\rm eff}^{-2}$. Assuming $L_\star \propto \dot{M}_{\rm w}^{0,5}v_{\rm w}^{0.5}$ \citep{Kobulnicky_2017_B}, and using Eqs.~(\ref{Eq:L_mec}) and (\ref{Eq:t_conv}), we estimate
\begin{equation}
    L_{\rm IC_\star} \sim L_{\rm NT} \,\left(\frac{t_{\rm conv}}{t_{\rm IC_\star}}\right) \propto 
    \begin{dcases}
    f(a)\,a^{-1}\,\dot{M}_{\rm w}\,v_{\rm w}\,n_{\rm ISM}^{0.5}\,\mu_{\rm ISM}^{0.5}\,V_\star    &{\rm RS}\\ \\
    a\,\dot{M}_{\rm w}\,v_{\rm w}\,n_{\rm ISM}^{0.5}\,\mu_{\rm ISM}^{0.5}\,V_\star    &{\rm FS}.
    \end{dcases}
\end{equation}

\noindent As a consequence, high energy emission depends mostly on the mass-loss rate and the velocities of the star and the wind. Massive stars moving with high velocities in a dense medium are the most promising $\gamma$-ray sources. However, as shown in the following section, even for these sources the expected fluxes are significantly below the sensitivity threshold of current $\gamma$-ray observatories.

\subsection{Spectral energy distribution}

\begin{figure}[ht]
    \centering
    \includegraphics[angle=270, width=0.36\textwidth]{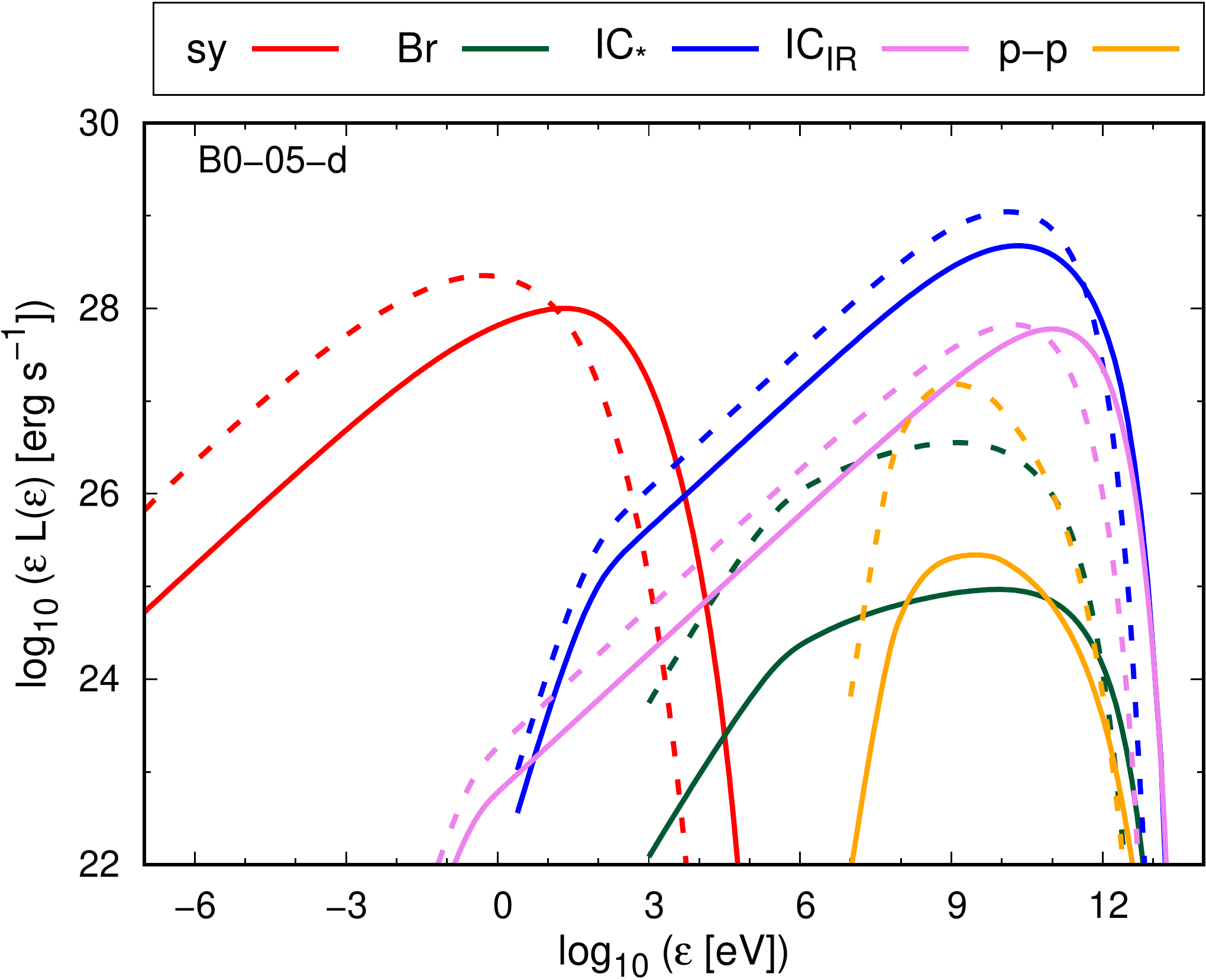}\\
    \includegraphics[angle=270, width=0.36\textwidth]{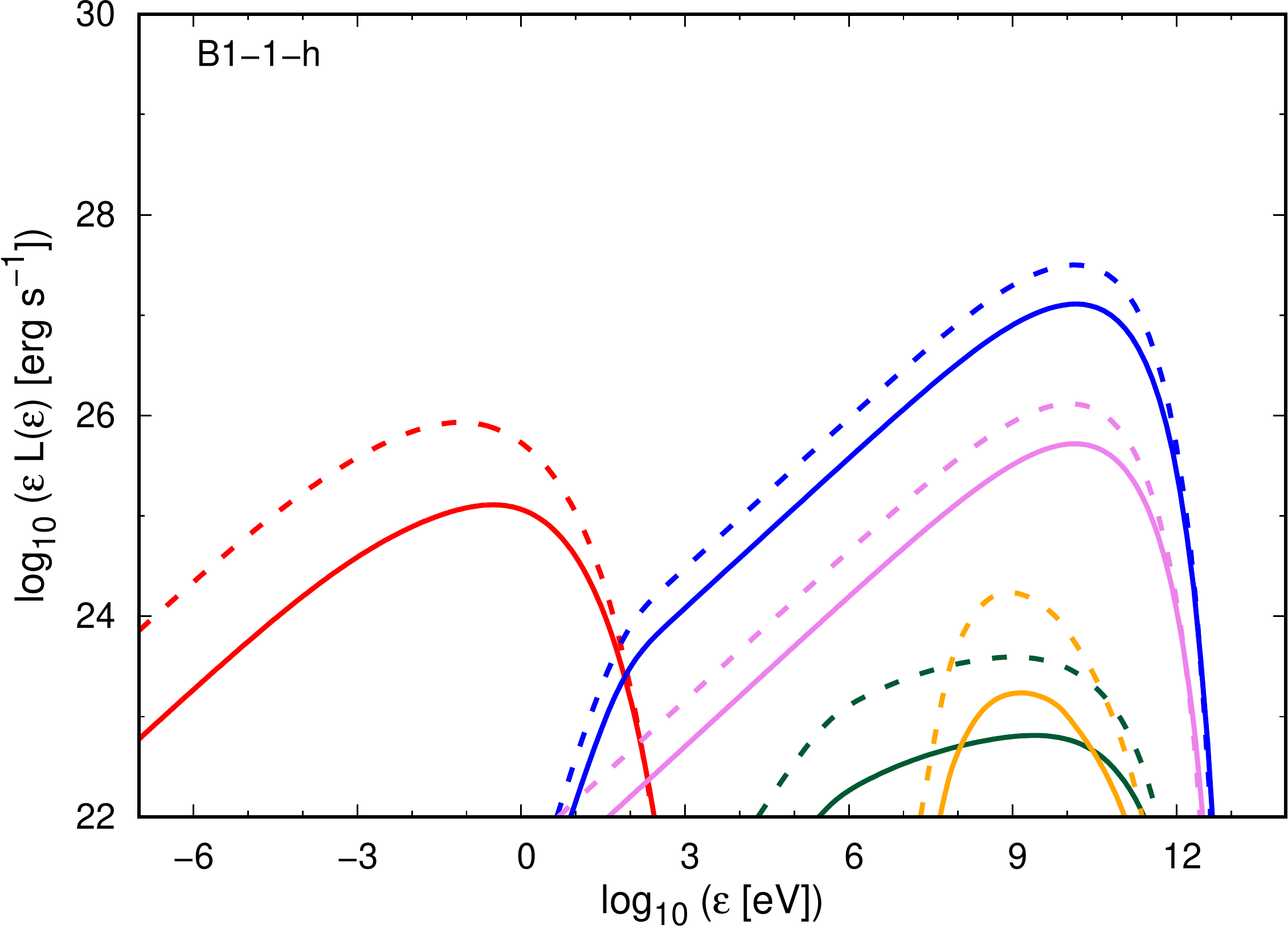}\\
    \includegraphics[angle=270, width=0.36\textwidth]{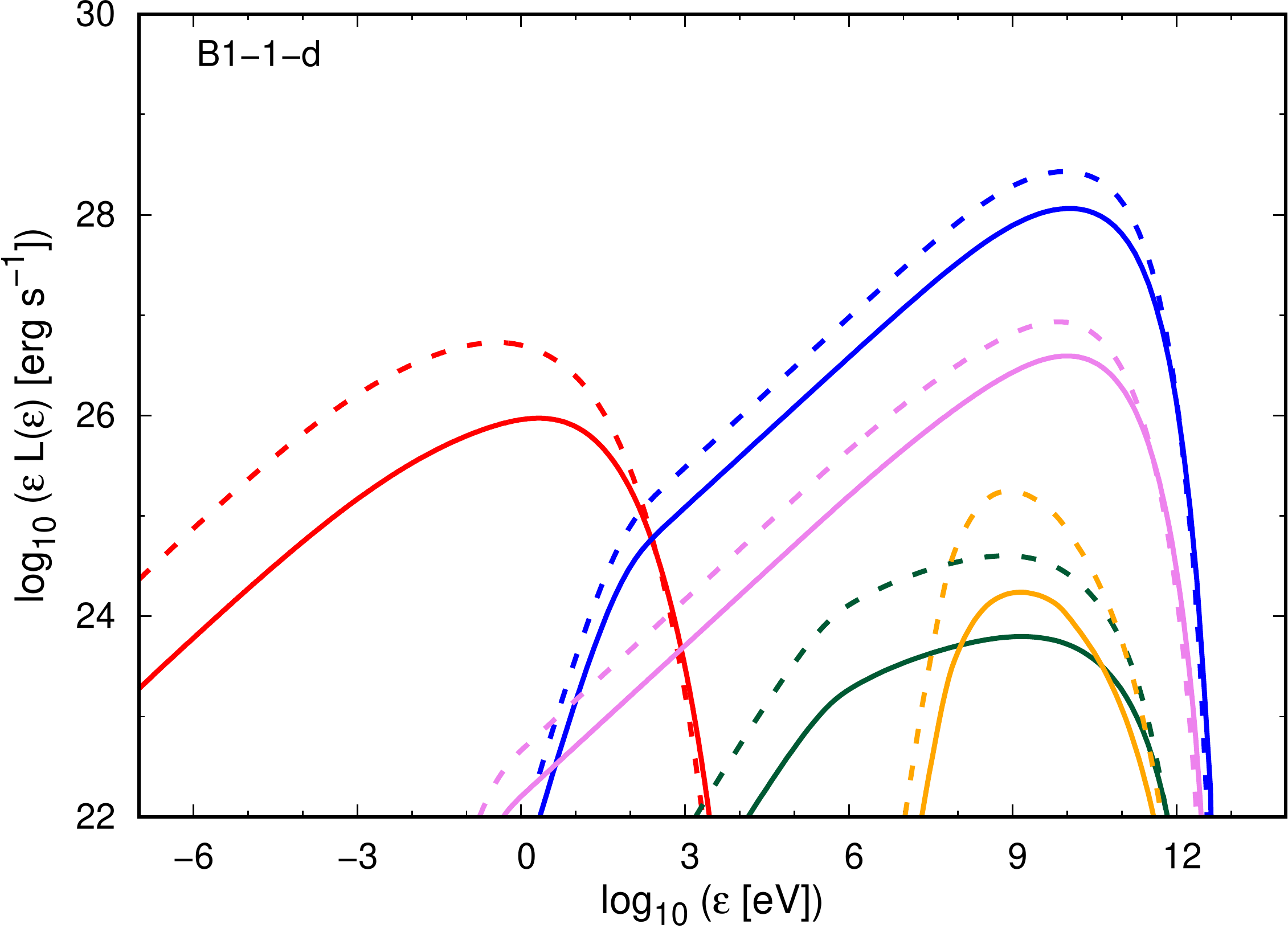}\\
    \includegraphics[angle=270, width=0.36\textwidth]{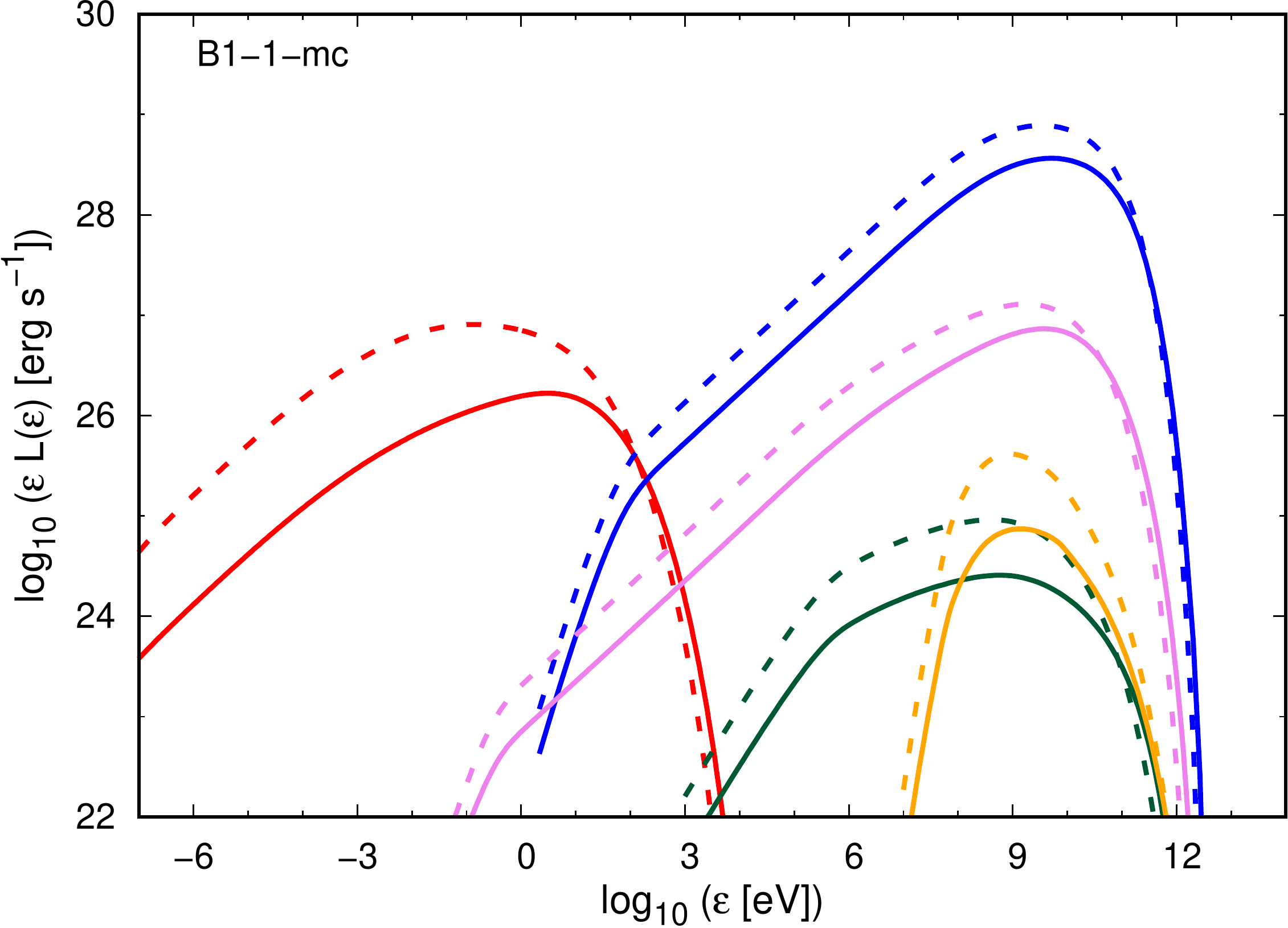}\\
    \includegraphics[angle=270, width=0.36\textwidth]{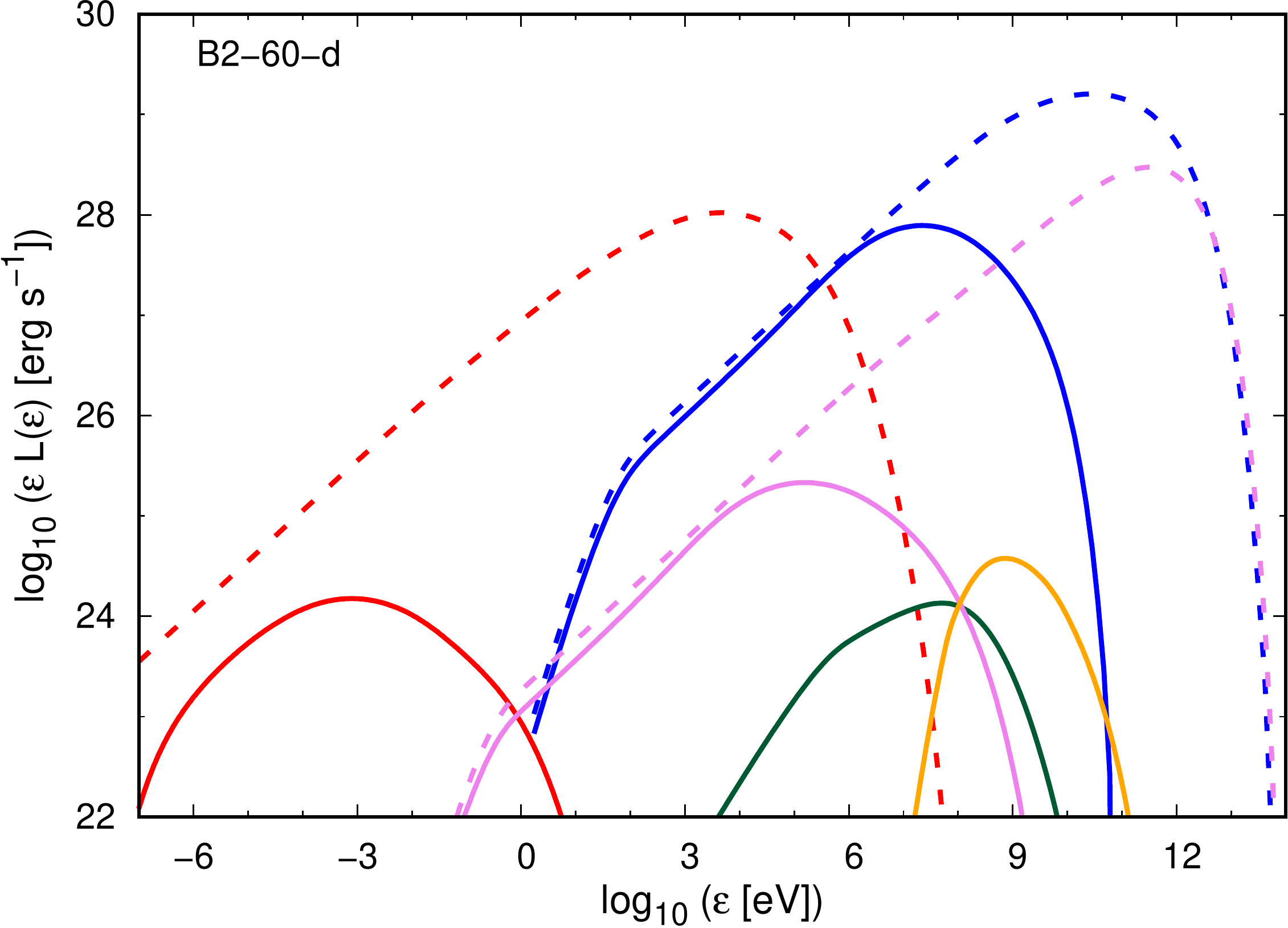}\\
    \caption[]{SEDs for the systems considering $f_\mathrm{NT} = 0.1$ and $\eta_B = 0.1$. Solid lines correspond to the RS, while dotted lines correspond to the FS.}
    \label{fig:SEDs}
\end{figure}

In Fig.~\ref{fig:SEDs} we show the SEDs obtained for all the systems studied. As expected, the NT spectrum is dominated by synchrotron emission in the radio band and IC$_\star$ emission in $\gamma$-rays. In the X-ray band, the IC$_\star$ component is usually dominant, although a non-negligible contribution from synchrotron radiation can also be expected for high magnetic fields, which can even be dominant in SRSs. On the other hand, relativistic Bremsstrahlung and hadronic NT emission are not relevant.

In all scenarios, the NT radiation of the FS is brighter than that of the RS. In addition, the system B0--05--d is the most luminous, in agreement with the discussion in Sec.~\ref{subsec:estimaciones}, where we showed that synchrotron emission depends mostly on the stellar wind parameters. On the other hand, the system B1--1--mc has the most luminous BS among systems with $V_\star = 1\,000~$km\,s$^{-1}$. Thus, as expected from Sec.~\ref{subsec:estimaciones}, a denser medium favours NT emission.

In Table~\ref{Tabla:flujos} we show the fluxes predicted at $\nu = 1.4$~GHz, assuming a distance to the star of 1~kpc. We conclude that a system with the characteristics of B0--05--d could be detected by the new generation of radio interferometers, such as the SKA \citep{Cassano2018} and the ngVLA \citep{McKinnon2019}, if it is at a distance $<3$~kpc. On the other hand, assuming $\eta_B = 1$ increases the fluxes by a factor of five, giving room for detection of slightly less favourable sources.

The SRS is the most luminous X-ray source among the systems considered. Given that electrons are accelerated up to energies $E_{\rm max} \sim 10$~TeV, the synchrotron spectrum reaches energies of $\epsilon \lesssim 1$~MeV in the FS. Nevertheless, we predict luminosities of $L_{\rm X} \sim 10^{28}$~erg s$^{-1}$ between 1 keV and 10 keV, undetectable by current 
instruments unless at a highly unlikely short distance. However, we highlight that by considering $\eta_{\rm B} = 1$ and that if a 50\% of $L_{\rm NT}$ went to electrons, this luminosity could increase by a factor of $\sim 200$, making it detectable by {\it Chandra} or future instruments like {\it Lynx}, if the system is at a distance $\lesssim 1$~kpc.

 Finally, we note that IC emission reaches energies of $\epsilon \gtrsim 10$~TeV in the FS of the system B2--60--d, for which IC with both the stellar and the IR field occur in the K-N regime. Nonetheless, the predicted $\gamma$-ray radiation is undetectable with present or forthcoming instrumentation in all systems. The fluxes predicted are at least two orders of magnitude below the detection threshold of CTA and \textit{Fermi}-LAT, even assuming distances of 1~kpc to the source \citep{Bruel2018, Maier2019}.


\begin{table}[t]
\centering
\caption[]{Energy fluxes predicted at $\nu = 1.4$~GHz assuming a distance of 1~kpc to the stars.}
\begin{tabular}{c| c c c| c c c}
\hline \hline
     & \multicolumn{6}{c}{S$_\mathrm{1.4\,GHz}$ [$\mu$Jy]} \\ \cline{2-7}
     & \multicolumn{3}{c|}{$\eta_B = 0.1$} & \multicolumn{3}{c}{$\eta_B = 1$} \\ \cline{2-7}
Scenario      & RS    & FS       & Total    & RS     & FS        & Total    \\
\hline
B0--05--d     & 24.3  & 317.4    & 341.7    & 132.5  & 1784    & 1916.5   \\          
B1--1--h      & 0.3  & 3.0     & 3.3     & 1.4   & 16.6      & 18.0   \\         
B1--1--d      & 0.9  & 10.9     & 11.8      & 4.6   & 56.8     & 61.4   \\            
B1--1--mc     & 1.9  & 24.0    & 25.9     & 10.2   & 124.6     & 134.8   \\            
B2--60--d     & 0.3  & 1.6     & 1.9     & 1.4   & 8.6     & 10.0   \\          
\hline
\end{tabular}
\label{Tabla:flujos}
\end{table}



\section{Conclusions} \label{sec:conclusions}

We investigated the shocks produced by massive stars that move with hypersonic or semi-relativistic velocities with respect to their surrounding medium. We introduced a refined model for calculating the emission from stellar BSs, especially relevant for systems moving with very high velocities ($V_\star > 300$~km\,s$^{-1}$). Our results show that in these systems both the RS and the FS are adiabatic and hypersonic, making them promising NT particle accelerators, potentially contributing with a $\sim 0.1$\% component to the galactic cosmic rays at energies $\lesssim 1$~TeV. Protons and electrons are accelerated in these systems up to energies $\gtrsim 1$~TeV. Nonetheless, the detection of NT emission associated with these relativistic particles is likely to remain elusive, as the predicted fluxes are too faint for current observatories. We suggest that, in the near future, the most promising observational breakthrough could be achieved by the next generation of interferometers operating at low radio frequencies, which can potentially detect the NT emission produced by BSs of early-type HVSs. We also note that under rather optimistic conditions for the production of leptonic radiation (i.e. $\eta_{\rm B} = 1$ and with a $\sim 5$\% of the available energy injected into non-thermal electrons) sources within 1~kpc from Earth might be detectable in X-rays with future instruments.

\begin{acknowledgements}
V.B-R. \& G.E.R. acknowledge financial support from the \emph{State Agency for Research} of the Spanish Ministry of Science and Innovation under grant PID2019-105510GB-C31 and through the \emph{Unit of Excellence Mar\'ia de Maeztu 2020-2023} award to the Institute of Cosmos Sciences (CEX2019-000918-M). V.B-R. is also supported by the Catalan DEC grant 2017 SGR 643, and is Correspondent Researcher of CONICET, Argentina, at the IAR.      
\end{acknowledgements}

%
%

\bibliographystyle{aa} 
\bibliography{bibliography} 


\begin{appendix}

\section{Comparison of the hydrodynamical model}\label{app:new_old}

\begin{figure}
    \centering
    \includegraphics[angle=90, width=0.8\linewidth, angle=270]{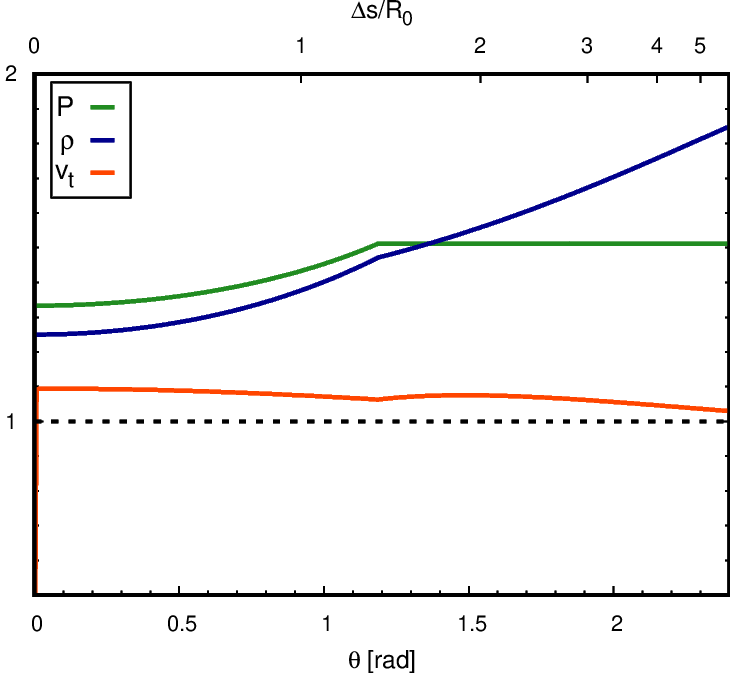}
    \caption{Ratio between the values given by the hydrodynamic prescription used in this work and the values using Rankine-Hugoniot jump conditions for strong shocks for the RS. The green, blue, and orange lines correspond to the pressure, density, and speed of the shocked fluid, respectively. The black dotted line at unity is marked as a reference.}
    \label{fig:new_vs_old}
\end{figure}

We compare the values of the pressure, density, and speed of the shocked fluid obtained with the prescriptions used in this work (Sec.~\ref{subsec:Hydro}) with those obtained by considering Rankine-Hugoniot jump conditions. We summarise the results in Fig.~\ref{fig:new_vs_old}, which shows the ratio between these quantities calculated with each formalism for different angles $\theta$. Both prescriptions yield very similar speeds for the shocked gas, although the prescription used in this work yield greater values for the remaining thermodynamic quantities. The discrepancy on the pressure increases with the angle $\theta$ until $\theta \sim 60\degree$, while the discrepancy on the density keeps increasing with $\theta$ and it reaches a difference of a factor two at $\theta \sim 140\degree$.


\end{appendix}

\end{document}